\documentclass[12pt]{iopart}
\usepackage{ifpdf}
    \ifpdf
    \usepackage[pdftex]{graphicx}  
      \usepackage[pdftex]{hyperref}
    \else
      \usepackage[dvips]{graphicx}  
      \newcommand{\href}[2]{#2}
    \fi
\usepackage{pslatex}
\usepackage{mathptm}    
\usepackage{mathptmx}   
\usepackage{dcolumn}
\usepackage{hyperref}
\usepackage{color}
\usepackage{booktabs,tabulary}
\usepackage{soul}
\usepackage[export]{adjustbox}
\usepackage{caption}
\usepackage{epstopdf}
\usepackage{subcaption}
\usepackage{epstopdf}

\usepackage{amssymb} 
\usepackage{bm} 

\newcommand{\Z}{\mathbb{Z}}
\newcommand{\R}{\mathbb{R}}
\newcommand{\Q}{\mathbb{Q}}

\renewcommand{\d}[1]{{d#1}} 
\newcommand{\fn}[2]{\mathinner{#1\mathopen{\left(#2\right)}}} 

 \newcommand{\E}[1]{\left\langle#1\right\rangle}
\newcommand{\abs}[1]{\left\vert #1 \right\vert}
\newcommand{\ceil}[1]{\lceil #1 \rceil}
\newcommand{\floor}[1]{\lfloor #1 \rfloor}
\newcommand{\fra}[1]{\left\{ #1 \right\}}
\newcommand{\vect}[1]{\bm{#1}}  \newcommand{\unitvect}[1]{\hat{\bm{#1}}} 

\newcommand{\atan}[1]{\fn{\tan^{-1}}{#1}}

\newcommand{\set}[2]{\left\{#1 ~\middle\vert ~#2 \right\}}
\newcommand{\eqref}[1]{(\ref{#1})}
\newcommand{\pder}[2]{\mathop{\frac{\partial #1}{\partial #2}}}
\newcommand{\conFrac}[2]
{\frac{#1}{\displaystyle #2}}

\usepackage[colorinlistoftodos,shadow,textwidth=18mm]{todonotes}

\newenvironment{cond}
	{\left\{
    \begin{array}{l l}
    }
    { 
    \end{array} 
    \right.
    }

\newenvironment{matrix_re}[1]
	{\left(\begin{array}{#1} }{ \end{array}\right)}

\begin{document}

\title[Effect of Window Shape on the Local Number Variance]{Effect of Window Shape on the Detection of Hyperuniformity via the Local Number Variance}

\author{Jaeuk Kim$^1$ and Salvatore Torquato$^{1,2,3,4}$}
\address{$^1$ Department of Physics, Princeton University, Princeton, New Jersey 08544, USA}
\address{$^2$ Department of Chemistry, Princeton University, Princeton, New Jersey 08544, USA}
\address{$^3$ Princeton Institute for the Science and Technology of Materials, Princeton University, Princeton, New Jersey 08544, USA}
\address{$^4$ Program in Applied and Computational Mathematics, Princeton University, Princeton, New Jersey 08544, USA}

\ead{torquato@princeton.edu}
\vspace{10pt}
\date{\today}

\begin{abstract}
Hyperuniform many-particle systems in $d$-dimensional space $\R^d$, which includes crystals, quasicrystals, and some exotic disordered systems, are characterized by an anomalous suppression of density fluctuations at large length scales such that the local number variance within a ``spherical" observation window grows slower than the window volume.
In usual circumstances, this direct-space condition is equivalent to the Fourier-space hyperuniformity condition that the structure factor vanishes as the wavenumber goes to zero.
In this paper, we comprehensively study the  effect of aspherical window shapes with characteristic size $L$ on the direct-space condition for hyperuniform systems.
For lattices, we demonstrate that the variance growth rate can depend on the shape as well as the orientation of the windows, and in some cases, the growth rate can be faster than the window volume (i.e., $L^d$), which may lead one to falsely conclude that the system is non-hyperuniform solely according to the direct-space condition.
We begin by numerically investigating the variance of two-dimensional lattices using ``superdisk" windows, whose convex shapes continuously interpolate between circles ($p=1$) and squares ({\color{black}$p\to\infty$}), as prescribed by a deformation parameter $p$, when the superdisk symmetry axis is aligned with the lattice.
Subsequently, we analyze the variance for lattices as a function of the window orientation, especially for two-dimensional lattices using square windows (superdisk when {\color{black}$p\to\infty$}).
Based on this analysis, we explain the reason why the variance for $d=2$ can grow faster than the window area or even slower than the window perimeter (e.g., like $\ln(L)$).
We then study the generalized condition of the window orientation, under which the variance can grow as fast as or faster than $L^d$ (window volume), to the case of Bravais lattices and parallelepiped windows in $\R^d$.
In the case of isotropic disordered hyperuniform systems, we prove that the large-$L$ asymptotic behavior of the variance is independent of the window shape for convex windows.
We conclude that the orientationally-averaged variance, instead of the conventional one using windows with a fixed orientation, can be used to resolve the window-shape dependence {\color{black}of} the direct-space hyperuniformity condition.
We suggest a new direct-space hyperuniformity condition that is valid for any convex window.
The analysis on the window orientations demonstrates an example of physical systems exhibiting commensurate-incommensurate transitions and is closely related to problems in number theory (e.g., Diophantine approximation and Gauss' circle problem) and discrepancy theory.
\end{abstract}
\vspace{2pc}
\submitto{\JSTAT}

\section{Introduction}\label{sec:intro}
A hyperuniform state matter is characterized by an anomalous suppression of density fluctuations at large length scales \cite{HypU_rev, two-phase_Zachary, HU_generalizations}.
The hyperuniformity concept provides a unified way to categorize crystals, quasicrystals, and certain exotic disordered systems \cite{HypU_rev, two-phase_Zachary,eg_Zachery_MRJ}. 
Disordered hyperuniform states lie between a crystal and liquid: they behave like perfect crystals in the manner in which they suppress large-scale density fluctuations and yet, like liquids and glasses, are statistically isotropic without Bragg peaks.
In this sense, disordered hyperuniform systems have a hidden order on large length scales, which endows them with novel physical properties \cite{eg_Jiao_AvainPhotoreceptor, app_PBG_Florescu1, app_PBG_Florescu, app_PBG_Man, eg_Leseur_dense_transparent}.
During the last decade, it has been discovered that these systems play a vital role in a number of problems across the physical, mathematical, and biological sciences.
Specifically, we now know that disordered hyperuniform materials can exist as both equilibrium and nonequilibrium phases, including maximally random jammed packings \cite{eg_Jiao_MRJ, eg_Berthier_MRJ,eg_Donev_MRJ},
Coulomb gas \cite{eg_Lebowitz_OCP, exact_OCP_Jancovici, SM_random_matrix_Dyson1},
certain fermionic and bosonic systems \cite{eg_Scardicchio_determinantalPointProcess, eg_Torquato_FermionicGas, eg_Feynman_liquidHelium},
liquids that freeze into degenerate disordered ground states \cite{eg_Marcotte_supercooled_nonequilibrium},
novel disordered photonic materials \cite{app_PBG_Florescu, app_PBG_Florescu1, app_PBG_Man},
spatial patterns of photoreceptors in avian retina \cite{eg_Jiao_AvainPhotoreceptor},
structure of bird feathers \cite{eg_feather_Noh},
highly excited states of ultracold gases \cite{eg_Lesanovsky_coldAtom},
terahertz quantum cascade lasers \cite{eg_Degl_quantumCascadeLaser},
driven nonequilibrium systems \cite{eg_Hexner_AbsorbingStates, eg_nonequilibrium_Scjrenk, eg_Jack_nonEquilibrium},
transparent dense disordered materials \cite{eg_Leseur_dense_transparent}, and number theory \cite{SM_random_matrix_Dyson1, Math_RiemannZeta_montgomery}.

\begin{figure}[htp]
\begin{center}
\includegraphics[width = 0.7 \textwidth]{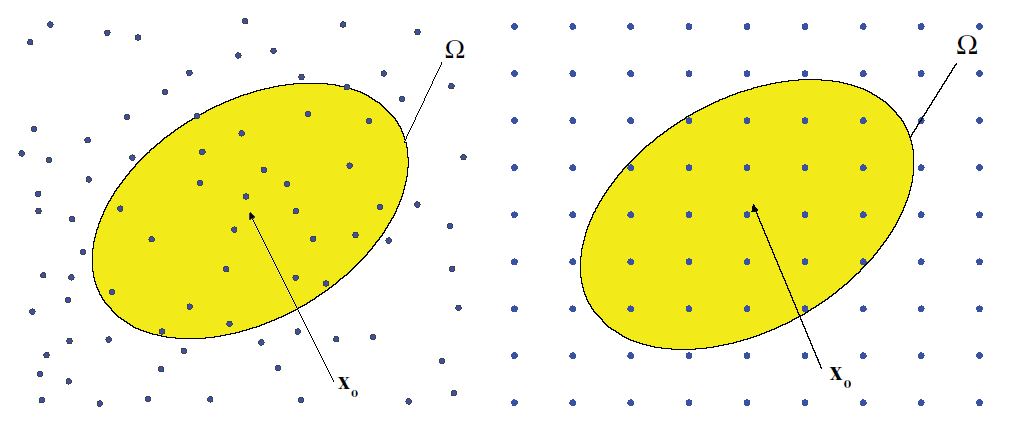}
\end{center}
\caption{(color online) Schematics indicating an observation window $\Omega$ and its centroid $\vect{x}_0$ for a disordered point pattern (left) and a periodic one (right), as adapted from \cite{HypU_rev}.\label{fig:sch_number variances}}
\end{figure}
Consider a point process in $d$-dimensional Euclidean space $\R^d$ and let $\fn{N}{\vect{R},\vect{x}_0}$ denote the number of points contained in a $d$-dimensional window $\Omega$, the shape of which is characterized by $\vect{R}$, and where $\vect{x}_0$ denotes the position of the centroid; see figure \ref{fig:sch_number variances}. 
Density fluctuations can be quantified by $\fn{\sigma_N ^2}{\vect{R}}$, i.e., the variance in $\fn{N}{\vect{R},\vect{x}_0}$ over either the ensemble of the point process or the centroidal position $\vect{x}_0$ of the window for a single realization of the point process.
The quantity $\fn{\sigma_N ^2}{\vect{R}}$ is directly related to the pair statistics of the point process and the geometry of the window $\Omega$ in the following way \cite{HypU_rev,two-phase_Zachary,HU_generalizations}:
\begin{equation}\label{eq:intro_sigma_n}
\fn{\sigma_N ^2}{\vect{R}} =\rho \fn{v_1}{\vect{R}} \left[1+\rho\int_{\R^d}  \fn{h}{\vect{r}}\fn{\alpha_2}{\vect{r};\vect{R}} \d{\vect{r}}  \right], 
\end{equation}
where $\rho$ is the number density of the point process, $\fn{v_1}{\vect{R}}$ is the volume of $\Omega$, and $\fn{\alpha_2}{\vect{r};\vect{R}} $ denotes the scaled intersection volume of $\Omega$, defined in \sref{sec:background_nv.HU}.
Here, $\fn{h}{\vect{r}}$ denotes the total correlation function of the point process, as defined in \sref{background1}.
Using the relation \eqref{eq:intro_sigma_n} and Parseval's theorem, one immediately obtains the Fourier representation of the local number variance \cite{HypU_rev}:
\begin{equation}\label{eq:intro_sigma_f}
\fn{\sigma_N ^2}{\vect{R}} =\frac{\rho \fn{v_1}{\vect{R}}}{(2\pi)^d} \int_{\R^d} \fn{S}{\vect{k}} \fn{\tilde{\alpha}_2}{\vect{k};\vect{R}} \d{\vect{k}},
\end{equation}
where $\fn{S}{\vect{k}}$ is the structure factor of the point process and $\fn{\tilde{\alpha}_2}{\vect{k};\vect{R}}$ represents the Fourier transform of $\fn{\alpha_2}{\vect{r};\vect{R}}$.

For a Poisson point process, $\fn{h}{\vect{r}} = 0$ for all $\vect{r}$, and hence \eqref{eq:intro_sigma_n} yields that the number variance grows as fast as the window volume $\fn{v_1}{\vect{R}}$.
This volume-like growth of $\fn{\sigma_N ^2}{\vect{R}}$ is typical of most disordered systems, including liquids and structural glasses \cite{HypU_rev, eg_Marcotte_supercooled_nonequilibrium, eg_non_equilibrium_Adam}.
A hyperuniform \cite{HypU_rev} (also known as ``superhomogeneous" \cite{eg_Gaberielli_UnivsereStructure}) point process is defined by the following infinite-wavelength behavior of the structure factor:
\begin{equation}\label{HU_def}
\lim_{\abs{\vect{k}} \to 0} \fn{S}{\vect{k}} = 0,
\end{equation}
which we call the Fourier-space hyperuniformity condition.
The use of definition \eqref{HU_def} in the relation \eqref{eq:intro_sigma_f} for spherical windows (and some aspherical windows with sufficiently smooth boundaries) implies that  hyperuniform point processes have vanishing {\it normalized} density fluctuations at large length scales as specified by \cite{HypU_rev, HU_generalizations}:
\begin{equation}\label{HU_criterion}
\lim_{\fn{v_1}{\vect{R}} \to \infty} \frac{\fn{\sigma_N ^2}{\vect{R}}}{\fn{v_1}{\vect{R}}} = 0,
\end{equation}
which is the usual \textit{direct-space} hyperuniformtiy condition.
Henceforth, we will call this the \textit{spherical-window} hyperuniform condition.
For spherical (and many apherical) windows, $\fn{\sigma_N ^2}{\vect{R}}$ for a hyperuniform point process has a growth rate that varies between the window surface area $\fn{s_1}{\vect{R}}$ and the window volume $\fn{v_1}{\vect{R}}$ in the large-window limit.
The variance for perfect crystals and a large class of quasicrystals grows asymptotically like $\fn{s_1}{\vect{R}}$.

The hyperuniformity concept has been extended to two-phase heterogeneous media \cite{two-phase_Zachary, HU_generalizations}. 
Here, one needs to use the spectral density $\fn{\tilde{\chi}_{_V}}{\vect{k}}$ associated with the appropriate two-point probability function and the local volume-fraction variance $\fn{\sigma_V ^2}{\vect{R}}$. 
Then, the Fourier-space hyperuniformity condition is
\begin{equation}\label{intro_HU_def_2phases}
\lim_{\abs{\vect{k}} \to 0} \fn{\tilde{\chi}_{_V}}{\vect{k}} = 0,
\end{equation}
and equivalently, the corresponding \textit{spherical-window} condition is
\begin{equation}\label{intro_HU_cond_2phases}
\lim_{\fn{v_1}{\vect{R}}\to \infty} \fn{v_1}{\vect{R}}\fn{\sigma_{V} ^2}{\vect{R}} = 0.
\end{equation}
Hyperuniform systems can be identified through small-angle scattering experiments \cite{app_elec_Hejna,app_elec_Xie}, yielding either the structure factor $\fn{S}{\vect{k}}$ or the spectral density $\fn{\tilde{\chi}_{_V}}{\vect{k}}$.
However, for some systems, e.g., colloidal suspensions \cite{Drey_diagnose, eg_Joost_Periodic_Collidal_exp}, nano-copolymers \cite{eg_Zito_copolymer} and simulations \cite{eg_Hexner_AbsorbingStates}, scattering experiments may not be available.
In such instances, one can measure the local number variance $\fn{\sigma_N^2}{\vect{R}}$ in direct space via microscopy to ascertain hyperuniformity of these systems \cite{Drey_diagnose}. 
In this paper, we focus on hyperuniform point processes.

Previous theoretical investigations on the number variance of hyperuniform systems have primarily focused on spherical windows.
By constrast, little attention has been paid to the analysis of aspherical windows, except for a few investigations \cite{Math_Kendall_1, Math_Beck_rect, volumeFraction_periodic_Quintanilla, Anisotropy_Zachary}, including Beck's study of the use of rectangular windows that have a fixed height and a special orientation to analyze the square lattice \cite{Math_Beck_rect}.
He showed that in this case, the variance can grow slower than the window surface area.
Zachary \etal \cite{Anisotropy_Zachary} studied the two-dimensional checkerboard model and square lattice decorated by identical squares, and showed that their volume-fraction variances decrease as slow as the inverse of the window volume, i.e., $\fn{v_1}{\vect{R}}^{-1}$, despite the fact that these systems are hyperuniform (see \eqref{intro_HU_cond_2phases}).
In summary, we see that for certain window shapes, the \textit{spherical-window} hyperuniformity criterion alone may lead one to falsely conclude that a hyperuniform system (as identified via the Fourier-space condition) is non-hyperuniform.

Thus, the overall objective of this paper is to understand quantitatively the effect of aspherical window shapes on the asymptotic growth rate of the variance, and to resolve the possible inconsistencies that may arise with respect to the spherical-window condition \eqref{HU_criterion}.
It is noteworthy that the study of $\fn{\sigma_N^2}{\vect{R}}$ for aspherical windows is an interesting problem in physics and mathematics in its own right.
For example, it has been shown that finding a window-shape that minimizes $\fn{\sigma_N ^2}{\vect{R}}$ of a point process is equivalent to designing a finite-ranged repulsive pair potential that leads the point process to be the ground state \cite{HypU_rev}.
For periodic systems and regular polyhedral windows, we will show that the window orientation with respect to the system significantly affects the large-window asymptotic behavior of $\fn{\sigma_N ^2}{\vect{R}}$ (\sref{sec:lattice}). 
This is also one of many physical examples \cite{aubry_transition, Aubrey_transition_atomicChain, Hofstadter_Butterfly,com-incom_graphene_BN_1,com-incom_graphene_BN_2,com-incom_graphene_BN_3}, in which (in)commensurability of a certain parameter plays a crucial role in the physical properties, e.g., friction coefficient \cite{Aubrey_transition_atomicChain} and Hall conductivity \cite{com-incom_graphene_BN_2}.
It is interesting to note that the analysis of this orientational dependence under the incommensurate condition (\sref{sec:irrational angles}) is closely related to Diophantine analysis in number theory \cite{Prob_Diophantine_Beck} and discrepancy theory \cite{Math_Beck_rect, Prob_Diophantine_Beck}.

\begin{figure}[htp]
\begin{center}
\includegraphics[width = 0.9\textwidth]{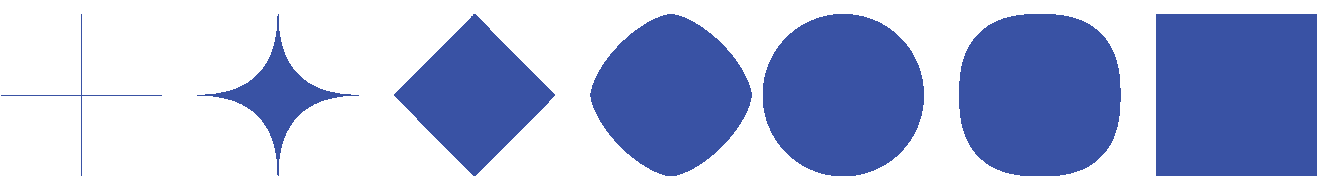}
\end{center}
\caption{Illustration of superdisk shapes for several values of the deformation parameters $p$. From the left to the right $p=0,~0.25,~0.5,~0.75,~1,~1.25,~\infty$.\label{fig:schematics_superdisk}}
\end{figure}

We begin by studying the variance for the lattices using aspherical windows that have a fixed orientation with respect to the lattices.
For this purpose, we numerically investigate the variance for the square lattice using the ``superdisk" windows.
A superdisk is the two-dimensional version of the versatile superball in $d$-dimensional Euclidean space $\R^d$, whose shape is defined by
\begin{equation}\label{intro:superdisk}
\abs{x_1}^{2p}+\abs{x_2}^{2p} +\cdots +\abs{x_d}^{2p} = L^{2p},
\end{equation}
where a positive real number $p$ is called \textit{deformation parameter} and $L$ is called the characteristic length scale.
If the parameter $p$ is smaller than 0.5, a superdisk is concave, and it interpolates smoothly between a cross ($p=0$) and a perfect square ($p=0.5$).
On the other hand, a superdisk with $p \geq 0.5$ is convex and is continuously transformed from a square ($p=0.5$) to the circle ($p=1$) and to a square of side length $2L$ ({\color{black}$p\to\infty$}), as shown in figure \ref{fig:schematics_superdisk}. 
Considering the case of $p\geq 1$, we show that the asymptotic behavior of the cumulative moving average of the variance has the power-law form, i.e., $\fn{\overline{\sigma_N ^2}}{L} \sim L^\gamma$ as $L \to \infty$.
We numerically demonstrate that the exponent $\gamma$ increases continuously from $1$ to $2$ as the window shape becomes closer to the perfect square, i.e., $p$ tends to $\infty$.
When the window is a perfect square ({\color{black}$p\to\infty$}), $\gamma=2$, the value of which might lead one to falsely conclude that the square lattice is not hyperuniform, since it conflicts with the spherical-window condition \eqref{HU_criterion}.
We say that for a $d$-dimensional hyperuniform system, the growth rate of the variance, as determined from the equation \eqref{eq:intro_sigma_n}, is ``anomalously" large whenever the exponent $\gamma\ge d$ because it is larger than what we expect from the ``spherical-window" condition \eqref{HU_criterion}.

Subsequently, we investigate the variance for the $d$-dimensional cubic (or hypercubic) lattice using hypercubic windows (superball when {\color{black}$p\to\infty$}) of side length $2L$ to understand the mathematical conditions under which the variance is anomalously large at large length scales.
When the windows are perfectly aligned with the lattice, we show that the variance grows like the square of the window surface area, i.e., $L^{2(d-1)}$ (see \ref{sec:generalizations of aligned}).
Surprisingly, this growth rate can be even faster than the window volume, $L^d$, and such large growth rate are typical of super-Poissonian point processes, such as systems at thermal critical points \cite{stat_mech_Text}.
For more detailed analysis, we numerically and analytically investigate the two-dimensional case partly because the square is amenable to exact analysis and partly because it is the most frequently used aspherical window to detect the hyperuniformity of a two-dimensional system in direct space.

We also demonstrate that the asymptotic growth rate of $\fn{\sigma_N ^2}{L,\theta}$ depends on the angle $\theta$ between the symmetry axes of the window and the lattice.
Importantly, we identify two classes of angles, at one of which, so-called rational angles, the variance grows ``anomalously largely", i.e., $\fn{\sigma_N ^2}{L,\theta}\sim L^2$  as $L\to \infty$.
We explain the origin of such an orientational dependence from two different points of view: the \textit{correlation} of density fluctuations concentrated in the vicinity of the window surface, and conditional convergence of the $2$nd moment ($d$-th moment in $\R^d$) of the total correlation function.
Based on the analysis, we generalize the concept of ``rational angles" for the square lattice and square window to Bravais lattices and parallelepiped windows in $\R^d$ (see \ref{sec:generalizations_2d}).
To discuss the conditional convergence of the second moment of total correlation function, we investigate this integral for the circular and square boundaries with Abelian summability method (see \ref{ConvergenceTest}).
In the case of $d$-dimensional isotropic disordered hyperuniform point processes, we prove that the asymptotic behavior of $\fn{\sigma_N ^2}{\vect{R}}$ is independent of the window shape if it is convex.

In addition, we suggest a new direct-space hyperuniformity condition \eqref{HU_direct new condition} using the orientationally-averaged local number variance $\E{\fn{\sigma_N ^2}{\vect{R}}}_O$.
We prove that for a $d$-dimensional anisotropic hyperuniform point process, corresponding to either a crystal or disordered one, $\E{\fn{\sigma_N ^2}{\vect{R}}}_O$ always exhibits the same asymptotic behavior for any convex window shape.
Then, we show that in the case of the square lattice and square windows, $\E{\fn{\sigma_N ^2}{L}}_O \sim L$, which is consistent with the case for circular windows.

In \sref{sec:background}, we describe basic definitions and mathematical equations to compute the variance $\fn{\sigma_N ^2}{\vect{R}}$ for any window shape.
We numerically compute the variance for the square lattice using superdisk windows of various shapes and show the relation between its asymptotic behavior and the deformation parameter $p$ in \sref{sec:superdisk}.
In \sref{sec:lattice}, we investigate the case of the two-dimensional square lattice with a square window.
In \sref{sec:disordered}, we consider disordered hyperuniform point processes and study the asymptotic behavior of their variance for convex windows, including square windows.
Orientationally-averaged variance is explained, and its asymptotic behavior is derived in \sref{sec:anistropic}.
Finally, we provide concluding remarks in \sref{sec:conclusion}.
A generalization of ``rational angles" to $d$-dimensional Bravais lattices and parallelepiped windows is presented in \ref{sec:generalizations_2d}.
Then, we carry out some example calculations for the cases of the square lattice and rectangular windows with a fixed aspect ratio, and the triangular lattice and square windows.
In \ref{sec:generalizations of aligned}, we show $\fn{\sigma_N ^2}{L}$ for $d$-dimensional hypercubic lattice with aligned hypercubic windows of side length $2L$.

\section{Background and definitions}\label{sec:background}
\subsection{Point processes}\label{background1}
Roughly speaking, a point process in $d$-dimensional Euclidean space $\R^d$ is a distribution of infinitely many points $\vect{r}_1,~\vect{r}_2,~\cdots,$ in $\R^d$. 
For statistically homogeneous point processes in $\R^d$ at a given number density $\rho$, $\rho^n \fn{g_n}{\vect{r}^n} \d{\vect{r}^n}$ represents the probability density for finding $n$ points at $\vect{r}^n = \vect{r}_1,~\vect{r}_2, ~\cdots,~\vect{r}_n$, and $\fn{g_n}{\vect{r}^n}$ is called the $n$-particle correlation function. 
The statistical homogeneity of a point process implies that $g_n$ is determined by only relative positions of $n$ particles, i.e., $\fn{g_n}{\vect{r}^n} = \fn{g_n}{\vect{r}_{21},\vect{r}_{31},\cdots,\vect{r}_{n1}}$ with $\vect{r}_{ij} \equiv \vect{r}_j -\vect{r}_i$ for $1\leq i\neq j \leq n$.

The pair correlation function $\fn{g_2}{\vect{r}}$ has a significant importance \cite{eg_Leseur_dense_transparent, Chandler_stat.mech, cornea_transparency_Farrell}.
In systems without long-range order, $\fn{g_2}{\vect{r}} \to 1$ as $\abs{\vect{r}} \to \infty$. 
Therefore, it is useful to introduce the total correlation function $\fn{h}{\vect{r}}$ defined as
 \begin{equation}
 \fn{h}{\vect{r}} \equiv \fn{g_2}{\vect{r}} -1,
 \end{equation} 
which decays to zero for large $\abs{\vect{r}}$ in the absence of long-range order. 
The structure factor $\fn{S}{\vect{k}}$ is related by Fourier transform of $\fn{h}{\vect{r}}$:
 \begin{equation}\label{background:structureFactore}
 \fn{S}{\vect{k}} = 1+\rho \fn{\tilde{h}}{\vect{k}}.
 \end{equation}

A (Bravais) lattice $\Lambda$ in $\R^d$ belongs to a special subgroup of point processes, which can be expressed as integer linear combinations of $d$ linearly independent vectors $\vect{a}_i$ for $i=1,~2,~\cdots,~d$, i.e.,
 \begin{equation}\label{background:lattice}
 \Lambda = \set{\vect{x} = \sum_{i=1}^d n_i \vect{a}_i \in \R^d}{\vect{n}=(n_1,~n_2,~\cdots,~n_d) \in \Z^d}.
 \end{equation}
 Every lattice $\Lambda$ has a reciprocal lattice $\Lambda^*$, which is a set of all reciprocal vectors $\vect{q}$ satisfying $\exp{\left(  i\vect{q}\cdot\vect{x}\right)} = 1$ for every $\vect{x} \in \Lambda$. 
 The structure factor $\fn{S}{\vect{k}; \Lambda}$ of the lattice $\Lambda$ is a sum of delta functions centered at each point in $\Lambda^*$ except for the origin: 
 \begin{equation}\label{structureFactor}
 \fn{S}{\vect{k};\Lambda} = \frac{(2\pi)^d}{v_c}\sum_{\vect{q} \in \Lambda^* \setminus \vect{0}} \fn{\delta}{\vect{k}-\vect{q}},
 \end{equation}
 where $1/v_c$ is the number density of lattice $\Lambda$, and $\fn{\delta}{\vect{k}}$ is the $d$-dimensional Dirac delta function. We will use the following definition of Fourier transform $\fn{\tilde{f}}{\vect{k}}$ and the inverse transform $\fn{f}{\vect{r}}$ (assuming their existence): 
 \begin{eqnarray}
 \fn{\tilde{f}}{\vect{k}} &= \int_{\R^d} \fn{f}{\vect{r}} e^{-i \vect{k} \cdot \vect{r}} \d{\vect{r}}, \label{fourier transform}\\
 \fn{f}{\vect{r}} &= \left(\frac{1}{2\pi}  \right)^d \int _{\R^d} \fn{\tilde{f}}{\vect{k}}e^{i \vect{k}\cdot\vect{r}} \d{\vect{k}}. \label{inverse fourier transform}
 \end{eqnarray}
For radially symmetric functions, i.e., $\fn{f}{\vect{r}} = \fn{f}{\abs{\vect{r}}}$ and $\fn{\tilde{f}}{\vect{k}} =\fn{\tilde{f}}{\abs{\vect{k}}}$, the Fourier and inverse Fourier transform can be expressed as
 \begin{eqnarray}
 \fn{\tilde{f}}{k} & = \left( 2\pi \right)^{d/2}\int_{0}^\infty r^{d-1}\fn{f}{r}  \frac{\fn{J_{d/2 -1}}{kr}}{\left(kr  \right)^{d/2 -1}} \d{r} \\
 \fn{f}{k} & = \left( \frac{1}{2\pi} \right)^{d/2}\int_{0}^\infty k^{d-1}\fn{\tilde{f}}{k}  \frac{\fn{J_{d/2 -1}}{kr}}{\left(kr  \right)^{d/2 -1}} \d{k},
 \end{eqnarray}
 where $\fn{J_\nu}{x}$ is the Bessel function of order $\nu$.
 
\subsection{Number variance and hyperuniformity}\label{sec:background_nv.HU}
Consider a statistically homogeneous point process at the number density $\rho$ in $d$-dimensional Euclidean space.
Using an observation window $\Omega$ whose shape and orientation are characterized by a set of parameters $\vect{R}$, one can obtain the exact expression for $\fn{\sigma_N ^2}{\vect{R}}$ in both the direct- and Fourier-space representations \cite{HypU_rev}:
\begin{eqnarray}
\fl\fn{\sigma_N ^2}{\vect{R}}\equiv\E{\fn{N^2}{\vect{R}}}-\E{\fn{N}{\vect{R}}}^2 &= \rho \fn{v_1}{\vect{R}}\left[1+\rho \int_{\R^d} \fn{h}{\vect{r}}\fn{\alpha_2}{\vect{r};\vect{R}} \d{\vect{r}} \right] \label{sigma2_direct} \\
\fl&=\frac{\rho \fn{v_1}{\vect{R}}}{\left(  2\pi\right)^d} \int \fn{S}{\vect{k}}\fn{\tilde{\alpha}_2}{\vect{k};\vect{R}} \d{\vect{k}}, \label{sigma2_Fourier}
\end{eqnarray}
where $\fn{S}{\vect{k}}$ is the structure factor, defined by \eqref{background:structureFactore}, and $\fn{\alpha_2}{\vect{r};\vect{R}}$ is the scaled intersection volume of two identical windows but separated by $\vect{r}$, i.e., $\Omega$ and $\Omega+\vect{r}$.
Note that both \eqref{sigma2_direct} and \eqref{sigma2_Fourier} represent the variance when windows have a fixed orientation with respect to the point process.
Here, $\fn{\alpha_2}{\vect{r};\vect{R}}$ is expressed as the convolution of the window indicator functions $\fn{w}{\vect{x};\vect{R}}$:
\begin{equation} \label{eq_alpha}
\fn{\alpha_2}{\vect{r};\vect{R}} \equiv \frac{\fn{v_2 ^\mathrm{int}}{\vect{r};\vect{R}}}{\fn{v_1}{\vect{R}}} = \frac{1}{\fn{v_1}{\vect{R}}}\int_{\R^d} \fn{w}{\vect{x};\vect{R}}\fn{w}{\vect{x}-\vect{r}; \vect{R}} \d{\vect{x}},
\end{equation}
where $\fn{v_1}{\vect{R}}$ is the window volume of $\Omega$.
Clearly, the volume integral of $\fn{\alpha_2}{\vect{r};\vect{R}}$ over $\R^d$ is equal to the window volume:
\begin{equation}\label{integral of alpha}
\fn{\tilde{\alpha}_2}{\vect{0};\vect{R}} = \int_{\R^d} \fn{\alpha_2}{\vect{r};\vect{R}} \d{\vect{r}} = \int_{\R^d} \fn{w}{\vect{r};\vect{R}} \d{\vect{r}} = \fn{v_1}{\vect{R}}.
\end{equation}
For a $d$-dimensional sphere of radius $R$, the analytical expression for $\fn{\alpha_2}{r;R}$ is well-known in any spatial dimension $d$  \cite{RHM_Torquato, LowBoundDensity_Tor_Stil}. 
The explicit expression for the Fourier transform of $\fn{\alpha_2}{r;R}$ is given by
\begin{equation}\label{F_intersectionV}
\fn{\tilde{\alpha}_2}{k;R} =\frac{\fn{\tilde{w}}{k;R}^2}{\fn{v_1}{R}} = 2^d \pi^{d/2}\fn{\Gamma}{1+d/2} \frac{\left[ \fn{J_{d/2}}{kR} \right]^2}{k^d}.
\end{equation}
Denoting $x\equiv r/(2R)$, $\fn{\alpha_2}{r;R}$ can be expressed in the series representation in terms of $x$ for $x<1$ \cite{LowBoundDensity_Tor_Stil}:
\begin{equation}\label{intersectionV_expansion}
\fn{\alpha_2}{r;R} = 1-\fn{c}{d}x +\fn{c}{d} \sum_{n=2}^\infty  \frac{(-1)^n \fn{\Gamma}{(d+1)/2}}{(2n-1)\fn{\Gamma}{n} \fn{\Gamma}{(d+3)/2-n}}x^{2n-1},
\end{equation}
where $\fn{c}{d} = 2 \fn{\Gamma}{d/2+1}/\left[	\fn{\Gamma}{(d+1)/2}\fn{\Gamma}{1/2} \right]$ and $\fn{\Gamma}{x}$ is the Gamma function.

For a given window $\Omega$ and a realization $X$ of a point process, 
$\E{\fn{N}{\vect{R}}}$ represents the ensemble average of the number of particles of $X$ within the window $\Omega$, $\fn{N}{\vect{R};X}$, over every realization $X$ of the point process.
The \textit{ergodic hypothesis} enables us to equate an ensemble average to a volume average in a single infinite realization $X$ \cite{RHM_Torquato},
and trivially $\E{\fn{N}{\vect{R}}} =\rho \fn{v_1}{\vect{R}}$. 
In practice, we implement both definitions to compute $\E{\fn{N}{\vect{R}}}$. 
The scaled intersection volume $\fn{\alpha_2}{\vect{r};\vect{R}}$ that appears in \eqref{sigma2_direct} can be interpreted as a monotonic repulsive pair potential energy with compact support, where $\vect{R}$ defines the interaction range.
Thus, the point configurations that globally minimize $\fn{\sigma_N ^2}{\vect{R}}$ in a fixed dimension $d$ correspond to the classical ground states associated with this repulsive pair potential \cite{HypU_rev}.
Due to the integrability requirement of \eqref{sigma2_direct}, the variance cannot increase faster than the square of window volume, $\left(\fn{v_1}{\vect{R}}  \right)^2$ \cite{two-phase_Zachary}.

The hyperuniformity concept has been recently generalized to treat systems with directionally-dependent structure factors, so-called ``\textit{directionally-hyperuniform}" systems \cite{HU_generalizations}.
In this paper, unless otherwise stated, hyperuniform systems refer solely to direction-independent hyperuniform point processes, defined by \eqref{HU_def}.

For spherical windows of radius $R$, substituting \eqref{intersectionV_expansion} into \eqref{sigma2_direct}, one can obtain \cite{HypU_rev}:
\begin{equation}\label{sigma2_expansion}
	\fn{\sigma_N ^2}{R} = \rho \fn{v_1}{1} \left[ \fn{A_N}{R} R^d + \fn{B_N}{R} R^{d-1} + \fn{o}{R^{d-1}}  \right],
\end{equation}
where $\fn{o}{x^a}$ represents all terms of order less than $x^a$, and coefficients $\fn{A_N}{R}$ and $\fn{B_N}{R}$ are given by
\begin{eqnarray}
\fn{A_N}{R} & = 1+\rho \int_{\abs{\vect{r}} <R} \fn{h}{\vect{r}}\d{\vect{r}} \label{sigma2_1stTerm} \\
\fn{B_N}{R} & = -\frac{\rho d \fn{\Gamma}{d/2}}{2\fn{\Gamma}{1/2} \fn{\Gamma}{(1+d)/2}} \int_{\abs{\vect{r}} <R } \fn{h}{\vect{r}}\abs{\vect{r}}\d{\vect{r}}, \label{sigma2_secondTerm}
\end{eqnarray}
where $\fn{B_N}{R}$ is essentially the $d$-th moment of the total correlation function $\fn{h}{r}$.
In the limit of $R\to \infty$, coefficients $A_N$ and $B_N$ are convergent for all periodic point patterns, a large class of quasicrystals, and some disordered point patterns whose total correlation functions $\fn{h}{r}$ decay to zero faster than $1/r^{d+1}$  \cite{HypU_rev,two-phase_Zachary,anomal_Zachary}.

Since $\lim_{R\to \infty} \fn{A_N}{R} =\lim_{\abs{\vect{k}}\to 0}\fn{S}{\vect{k}}$, a hyperuniform system has the vanishing scaled long-wavelength density fluctuations, i.e., $\lim_{R \to \infty} \fn{\sigma_N ^2}{R} /\fn{v_1}{R} =0$.
When the structure factor goes to zero with power-law form $\fn{S}{\vect{k}} \sim \abs{\vect{k}}^\alpha$ and $0<\alpha \leq 1$, the coefficient term $\fn{B_N}{R}$, defined by \eqref{sigma2_secondTerm}, asymptotically converges to a function of $R$ and hence,  \cite{HypU_rev,two-phase_Zachary,eg_Torquato_FermionicGas}
\begin{equation}\label{eq:scaling relation}
\fn{\sigma_N ^2}{R}\sim 
\begin{cond}
	R^{d-\alpha}, &0<\alpha <1 \\
	R^{d-1}\ln{R}, &\alpha = 1 \\
	R^{d-1}, &\alpha >1 
\end{cond}~~,(R\to \infty).
\end{equation}
For some systems, e.g., lattices, the associated variances oscillate around some global average behavior so that it can be difficult to obtain smooth asymptotic behaviors. 
In such cases, it is advantageous to use the cumulative moving average of the variance $\fn{\overline{\sigma_N ^2}}{R}$ \cite{HypU_rev,Math_Kendall_1}, defined as
\begin{equation}\label{def_cumulative moving average}
\fn{\overline{\sigma_N ^2}}{R} \equiv \frac{1}{R} \int_0 ^R \fn{\sigma_N ^2}{x} \d{x},
\end{equation}
to observe the asymptotic behavior.

\subsection{Basic quantities for a square window and square lattice}\label{sec:Square Lattice}
Consider a square lattice in $\R^2$ of unit lattice constant.
Its total correlation function can be expressed as 
\begin{equation}\label{h_square}
\fn{h}{\vect{r}} = \prod_{i=1}^2 \left[ \sum_{n=-\infty}^{\infty} \fn{\delta}{r_i -n} \right] -\fn{\delta}{\vect{r}} - 1,
\end{equation}
where $r_i$ is $i$-th component of position vector $\vect{r}$.
Its isotropic form is
\begin{equation}\label{h_square_radial}
\fn{h}{r} = \sum_{k=1}^\infty  \frac{Z_k}{2\pi r_k}\fn{\delta}{r -r_k} - 1,
\end{equation}
where $r_k$ is the radius of $k$-th shell, and $Z_k$ is the corresponding coordination number.
Both expressions immediately follow from the definition of the square lattice and pair correlation function $\fn{g_2}{\vect{r}}$, given in Section \eqref{background1}.
The structure factor forms another square lattice, excluding the lattice point at the origin:
\begin{equation}
\fl\fn{S}{\vect{k}} = \left(2\pi  \right)^2 \sum_{\vect{n} \in \Z^2 \setminus \vect{0}} \fn{\delta}{\vect{k} - 2\pi\vect{n}} = \left(2\pi  \right)^2\left[\prod_{i=1}^2 \left( \sum_{n=-\infty}^{\infty} \fn{\delta}{k_i -2\pi n} \right) -\fn{\delta}{\vect{k}}	\right].
\end{equation}

The scaled intersection volume of two identical, aligned square windows of side length $2L$ \cite{Anisotropy_Zachary}, whose centers are separated by $\vect{r}$, is given by 
\begin{equation}\label{alpha_cubic-d}
\fn{\alpha_2}{\vect{r}; L} = \prod_{i=1}^2 \left(1-\frac{\abs{r_i}}{2L}  \right)\fn{\Theta}{2L - \abs{r_i}},
\end{equation}
where $\fn{\Theta}{x}$ is the Heaviside step function, defined as
\begin{equation}
\fn{\Theta}{x} = \begin{cond}
				1, & x\geq 0\\
				0, & x<0
				\end{cond}.
\end{equation}
The Fourier transform of $\fn{\alpha_2}{\vect{r};L}$ is
\begin{equation}\label{alpha_cubic_f}
\fn{\tilde{\alpha}_2}{\vect{k};L} = \left( \frac{2}{L} \right)^2\prod_{i=1}^2 \left( \frac{\sin{\left( k_i L \right)}}{k_i} \right)^2 .
\end{equation}
We note in passing that $\fn{\tilde{\alpha}_2}{\vect{k};L}$ is the same as the intensity profile of Fraunhofer diffraction pattern through a square aperture of side length $2L$. From this analogy, we can know that ``brightest" spots of \eqref{alpha_cubic_f} lie on principal axes on which either $k_x$ or $k_y$ is zero. On those principal axes, $\fn{\tilde{\alpha}_2}{\vect{k};L}$ is expressed as 
\begin{equation}\label{Foureir_alpha_on axes}
\fn{\tilde{\alpha}_2}{(0,k_y);L} = 4 \left( \frac{\sin{\left( k_y L \right)}}{k_y} \right)^2,
\end{equation}
and the magnitudes of peaks are proportional to $L^2$.
 
\section{Variance for square lattice with superdisk windows}\label{sec:superdisk}

\begin{figure}[htp]
\begin{center}
\begin{subfigure}[b]{ 0.50 \textwidth}
\includegraphics[width = \textwidth]{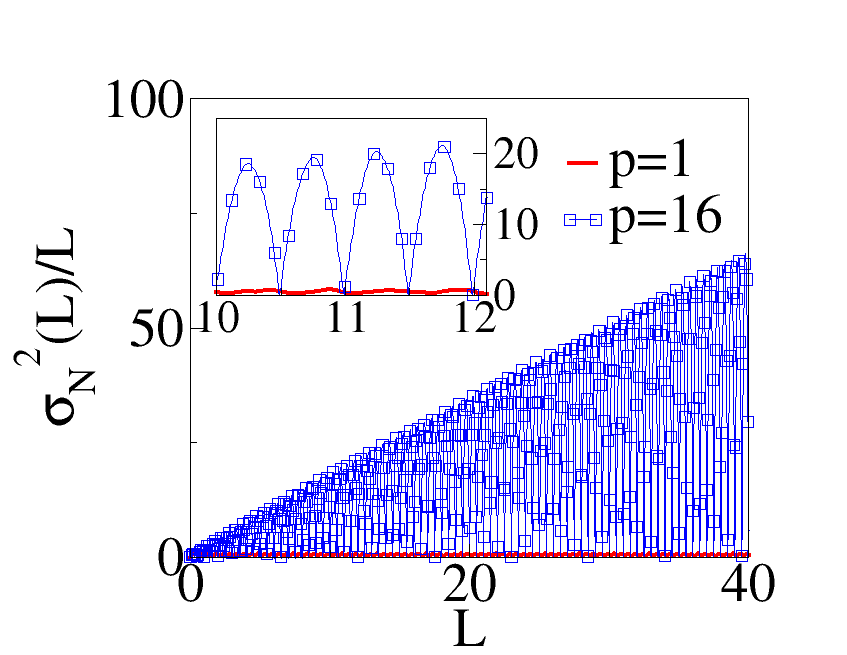}
\caption{}
\end{subfigure}
\begin{subfigure}[b]{ 0.48 \textwidth}
\includegraphics[width = \textwidth]{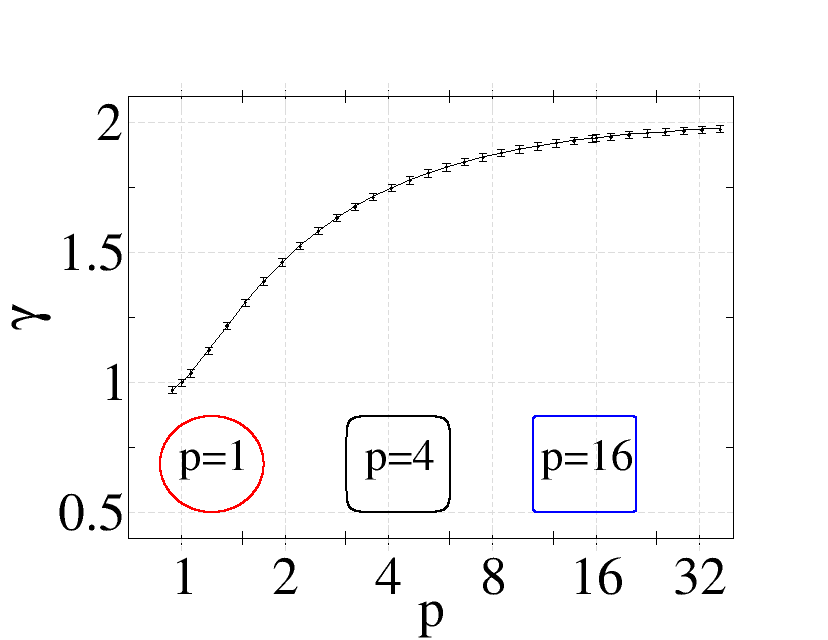}
\caption{\label{fig:superdisk}}
\end{subfigure}
\end{center}
\caption{The variance $\fn{\sigma_N ^2}{L}$ for the square lattice using superdisk windows {\color{black}obtained by the Monte Carlo calculations.}
(a) A plot of $\fn{\sigma_N ^2}{L}$ vs $L$ for the cases of $p=1,~16$. Note that the variance is divided by the characteristic length $L$. The inset is a magnification of the larger panel to show the case of $p=1$, which is otherwise not visible in the larger panel. 
(b) A semi-log plot of power exponent $\gamma$ of $\fn{\overline{\sigma_N ^2}}{L} \sim L^\gamma~(L\to \infty)$ vs the deformation parameter $p$.
{\color{black}To compute the exponent $\gamma$, we applied a linear regression to the log-log plot of $\fn{\overline{\sigma_N ^2}}{L}$, which was obtained by the Monte Carlo method up to $L=70$. }
The three inset figures show superdisks for certain deformation parameters.
 \label{fig:nv_superdisk_new}}
\end{figure}

The asymptotic expression for $\sigma_N ^2$ for non-circular windows has been intensively studied in $\Z^2$ (the square lattice) in various contexts, including the ``lattice-point counting problem" in number theory \cite{Math_Kendall_1,Math_Kendall_2, Math_planar_convex_body_LatticePts_Brandolini, Math_LatticePoints_DiophantineConditions_Alex}, discrete math \cite{Math_Beck_rect,Prob_Diophantine_Beck}, and stereology \cite{Matern1}.
It has been known that the asymptotic behavior of $\sigma_N^2$ can sensitively depend on the window shape \cite{Math_Kendall_1, Math_Kendall_GeometricProbabilty} as well as the orientation of the window \cite{Math_Beck_rect,Math_Rosen_rationalRectangle}. 
For instance, $\fn{\sigma_N ^2}{a} \sim a^{3/2}$ when the window is $\Omega=\set{(x,y)\in \R^2}{(xa)^2 +y^4 \leq a^4}$ \cite{Math_Kendall_1}, and $\fn{\sigma_N^2}{W, H} \sim W^2 + H^2$ when the window is a $W \times H$ rectangle whose sides are parallel to the principal axes of the lattice \cite{Math_Kendall_GeometricProbabilty}.
All of these asymptotic behaviors are different from the linear growth rate in $R$, which one expects using circular windows \cite{HypU_rev, Math_Kendall_1}.

In this section, to observe how the window shape can affect the growth rate of the variance, we measure the variance for the square lattice using a superdisk window with a fixed orientation.
Superdisks are two-dimensional figures whose shapes are described by the equation
\begin{equation}\label{eq:superdisk}
\abs{x}^{2p} +\abs{y}^{2p} = L^{2p},
\end{equation}
where $L$ is called the characteristic length scale, and $p$, also known as deformation parameter, is a positive real number.
Superdisks are ideal for our purpose to probe the effect of the window shape on the variance because one can generate a family of superdisks just by changing the parameter $p$.
When {\color{black}$p\to\infty$}, a superdisk is just a square of side length $2L$.
As $p$ decreases from $\infty$ to $1$, superdisks smoothly interpolates between the square ({\color{black}$p\to\infty$}) to the circle ($p=1$).
When $p<0.5$, the superdisk is concave and becomes a cross in the limit $p\to 0$.

Figure \ref{fig:nv_superdisk_new}(a) demonstrates the variances for square lattice using two different superdisks: one is a circular window ($p=1$) and another is a virtually square window ({\color{black}$p = 16$}).
As one can expect, $\fn{\overline{\sigma_N ^2}}{L} \sim L$ when $p=1$.
Importantly, when the window shape is a perfect square ({\color{black}$p\to\infty$}), one can obtain the closed expression for the variance by substituting \eqref{h_square} and \eqref{alpha_cubic-d} into \eqref{sigma2_direct}:
\begin{equation}\label{sigma2_0}
\fn{\sigma_N ^2}{L} = \fn{g}{2L} \left(2(2L)^2 +\fn{g}{2L} \right),
\end{equation}
where the function $\fn{g}{x}$ is defined as
 \begin{equation}\label{g-function}
 \fn{g}{x} \equiv \fra{x} \left( 1-\fra{x} \right),
 \end{equation} 
and $\fra{x}$ represents the fractional part of a positive real number $x$.
The cumulative moving average of \eqref{sigma2_0} is
\begin{equation}\label{sigma2_0_av}
\fn{\overline{\sigma_N ^2}}{L} \approx \frac{4}{9}L^2 +\fn{O}{L}   ~~(L\to \infty).
\end{equation}
Thus, as the parameter $p$ continuously increases from $1$ to $\infty$, the growth rate of the variance will vary from $L$ to $L^2$, i.e., \begin{equation}
\fn{\overline{\sigma_N ^2}}{L}\sim L^\gamma ~~(L\to \infty),
\end{equation}
where {\color{black}$\gamma$} increases from $1$ to $2$ (see figure \ref{fig:nv_superdisk_new}(b)).
This demonstrates that hyperuniform systems may exhibit anomalously large long-wavelength density fluctuations for aspherical windows, which is inconsistent with the ``spherical-window" hyperuniformity condition \eqref{HU_criterion}.

\section{Square-window variance for periodic point configurations}\label{sec:lattice}
In this section, we investigate the effect of the window orientations on the growth rate of the variance for the square lattice, $\Z^2$.
We use square windows because this shape is amenable to exact analysis for some quantities as well as is one limit of superdisk ({\color{black}$p\to\infty$}).
For this purpose, we denote by $\fn{\sigma_N ^2}{L;\theta}$ the variance for $\Z^2$ using square windows that have side length $2L$ and is rotated counterclockwise by an angle $\theta$ with respect to the lattice.
Here, it is sufficient to consider $\theta \in [0,\pi/4)$ due to 4-fold rotational symmetry and parity inversion symmetry of both the window and the lattice.
We identify two classes of angles, rational and irrational angles, according to the asymptotic behavior of $\fn{\sigma_N ^2}{L;\theta}$.
For rational angles, we derive the exact expression for $\fn{\sigma_N ^2}{L;\theta}$ and its asymptotic expression, and then compared these expressions with the Monte Carlo calculations.
For irrational angles, we compute the exact values for $\fn{\sigma_N ^2}{L;\theta}$ and conjecture its asymptotic behavior at a special set of irrational angles. 

We obtain expressions of $\fn{\sigma_N ^2}{L;\theta}$ in both direct and Fourier space representations.
Using \eqref{h_square} and \eqref{alpha_cubic-d}, the direct-space formula \eqref{sigma2_direct} yields
\begin{equation}\label{sigma2_direct_square}
\fn{\sigma_N ^2}{L;\theta} = \left(2L  \right)^2 \left[1-4L^2+  \sum_ {\vect{n}\in \Z^2 \setminus\vect{0}} \fn{\alpha_2}{R^T \vect{n} ; L} \right],
\end{equation}
where $\fn{\alpha_2}{\vect{r};L}$ is given in \eqref{alpha_cubic-d} and the rotation matrix $R$ is
\begin{equation}\label{RotationMatrix}
R = \begin{matrix_re}{cc}
	\cos{\theta} & -\sin{\theta} \\
	\sin{\theta} & \cos{\theta}
\end{matrix_re}.
\end{equation}
The summation in \eqref{sigma2_direct_square} 
can be written as
\begin{eqnarray}
\fl& \sum_ {\vect{n}\in \Z^d \setminus\vect{0}} \fn{\alpha_2}{R^T \vect{n} ; L}   =  4 \sum_{j=0}^{\fn{M}{L,\theta}} \left(n_2-n_1 +1 \right)\Bigg[1-\frac{\left( n_2+n_1 \right)\sin{\left(\frac{\pi}{4} - \theta  \right)} + 2j\sin{\left(\frac{\pi}{4}+\theta  \right)} }{2\sqrt{2}L}  \nonumber \\
\fl& + \left(j \left(n_1+n_2  \right) \frac{\cos{(2\theta)}}{8L^2}+\left( n_1 n_2 + \frac{1}{6}\left( n_2-n_1\right)\left( 2 (n_2-n_1)+1 \right)   -j^2 \right)\frac{\sin{(2\theta)}}{8L^2} \right) \Bigg], \label{directSpace_sum}
\end{eqnarray}
where $\fn{M}{L,\theta} = \floor{2\sqrt{2}L\cos{\left(\pi/4 -\theta  \right)}}$, $n_1$ and $n_2$ are abbreviations of $\fn{n_1}{j}$ and $\fn{n_2}{j}$, which are defined as
\begin{eqnarray}
\fn{n_1}{j} & = \max{\left\{\ceil{j \tan{\theta}} +1, \ceil{\cot{\theta}\left(j-2L\sec{\theta}  \right) }   \right\}} \label{LB of n}\\
\fn{n_2}{j} & = \min{\left\{ \floor{j \cot{\theta}}, \floor{\tan{\theta}\left( 2L\csc{\theta}  -j\right)} \right\}}, \label{UB of n}
\end{eqnarray}
where $\floor{x}$ means the largest integer less than or equal to $x$, and $\ceil{x}$ is the smallest integer larger than or equal to $x$.

The Fourier-space formula \eqref{sigma2_Fourier} yields
\begin{eqnarray}
\fn{\sigma_N ^2}{L ; \theta} & = \left(\frac{L}{\pi} 	\right)^d \int \fn{S}{\vect{k}} \fn{\tilde{\alpha}_2}{\vect{k};L} \d{\vect{k}} \label{sigma2_Fourier_square}  \\
&= \frac{1}{\pi^4} \sum_{\vect{n} \in \Z^2 \setminus \vect{0}} \left( \prod_{i=1}^2 \frac{\sin{^2 \left(2\pi \left[R^T \vect{n}  \right]_i L  \right)}}{\left[R^T \vect{n}  \right]_i ^2} \right), \label{sigma2_general_Foureir}
\end{eqnarray}
where $R$ is the rotation matrix given by \eqref{RotationMatrix}, and $\left[ \vect{x} \right]_i$ denotes the component of a vector $\vect{x}$ along the $i$-th principal axes of the square window.

\subsection{Rational angles}\label{sec:rational Angles}
An angle $\theta$ is called a \textit{rational angle} if its tangent is a rational number, i.e.,
\begin{equation}\label{def_rational angle}
\tan{\theta} \in \Q.
\end{equation}
In fact, the definition of rational angles can vary with the lattice and window (see \ref{sec:generalizations_2d}).
For a rational angle $\theta=\atan{n/m}$, where $n$ and $m$ are coprime integers, we can express the positions of Bragg peaks in \eqref{sigma2_general_Foureir} as those of the square lattice of lattice constant $L_0$ with $L_0$ basis points: 
\begin{equation}\label{NewLattice}
\fl \Lambda^* = \set{(i L_0 + k\cos{\theta} ,~j L_0 + k\sin{\theta})}{(i,j)\in \Z^2,~k=1,2,\cdots,L_0} \setminus\vect{0},
\end{equation}
where $L_0 = \sqrt{n^2 +m^2}$.
Substituting \eqref{NewLattice} into \eqref{sigma2_general_Foureir}, we obtain the expression for the variance
\begin{eqnarray}
\fl\fn{\sigma_N ^2}{L;\theta} & =
\sum_{k=1}^{{L_0}^2} \left[\sum_{i= -\infty}^{\infty} \frac{\sin{^2\left(2 \pi (i+x^{(k)})x  \right)}  }{\pi^2 (i +x^{(k)})^2 {L_0}^2} \right] \left[ \sum_{j= -\infty}^{\infty} \frac{\sin{^2\left(2 \pi (j+y^{(k)})x  \right)}  }{ \pi^2 (j +y^{(k)})^2 {L_0}^2} \right]  - \frac{16  x^4}{ {L_0} ^4} \label{sigma_inte0} \\
\fl	&= \frac{\fn{g}{2x}}{{L_0} ^4} \left(8x^2 +\fn{g}{2x} \right) +\frac{1}{{L_0} ^4}\sum_{k=1}^{{L_0}^2 -1} \fn{B}{x,x^{(k)}}\fn{B}{x,y^{(k)}}, \label{sigma_inte}
\end{eqnarray}
where $x={L_0}  L$, $x^{(k)} =\fra{nk /{L_0}^2} $, $y^{(k)} =\fra{mk /{L_0}^2} $, $\fra{x}$ is the fractional part of a real number $x$, $\fn{g}{x}$ is defined by \eqref{g-function}, and
\begin{eqnarray}
\fl\fn{B}{x,y} & \equiv \sum_{i= -\infty}^{\infty} \frac{\sin{^2\left(2 \pi (i+y)x  \right)}  }{ \pi^2 (i +y)^2 } \nonumber \\
\fl	&= \csc{^2 \left(\pi y  \right)} \Big( \sin{^2 \left( \pi y \floor{2x} \right)} +\sin{\left(  \pi y\right)} \sin{\left(\pi y \left(2\floor{2x}+1  \right)  \right)} \fra{2x}\Big).\label{B-function}
\end{eqnarray}
Note that the closed-form expression for $\fn{B}{x,y}$ in \eqref{B-function} is valid when $y \in \Q$, and this expression is equivalent to piecewise linear interpolation of $\csc{^2( \pi y)}\sin{^2 (2\pi yx)}$ whose data points are at $x=i/2$ for every integer $i$.
A proof of \eqref{B-function} {\color{black}is given in our Supplementary Data.}
At $\theta = 0$, \eqref{sigma_inte} {\color{black}becomes equal to} \eqref{sigma2_0}.
We note that relation \eqref{sigma2_0} was presented in \cite{Math_Kendall_GeometricProbabilty}, and relation \eqref{sigma_inte} was derived by Rosen \cite{Math_Rosen_rationalRectangle} for rectangular windows using a different approach.
Their derivations, however, highly relies on the geometry of the lattice and the window, and thus they are difficult to generalize to other lattices and other window shapes. 
On the other hand, it is straightforward to generalize the formula \eqref{sigma2_Fourier_square} to other lattices and other window shapes (see \ref{sec:generalizations_2d}).
Furthermore, relation \eqref{sigma_inte0} provides a useful insight to explain the origin of the anomalously large density fluctuations (see figure \ref{fig:FourierSpace interpretations}(a)).

\begin{figure}[htp]
		\centering
		\includegraphics[width = 0.75\textwidth]{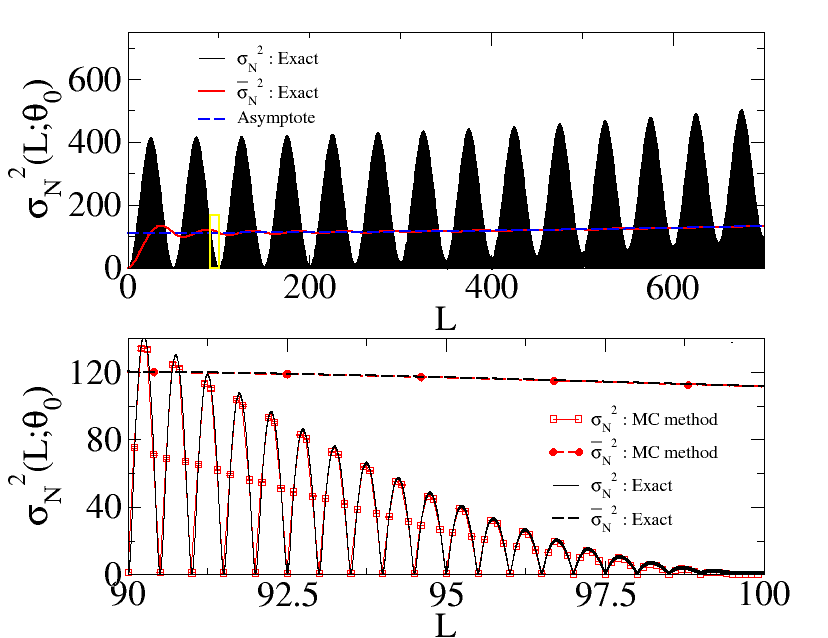}
		\caption{A plot of $\fn{\sigma_N ^2}{L;\theta_0}$ vs $L$, where $\theta_0 = \atan{1/100}$. It shows a typical behavior of variance at rational angles. Top panel shows the comparison between $\overline{\sigma_N ^2}$ and its asymptote \eqref{running_asymptote}.
		 The bottom panel is a magnification of the yellow-boxed region in the top panel. It demonstrates that the exact result by \eqref{sigma_inte} is consistent with the Monte Carlo calculations.  		
		 \label{fig:1/100 graph}}		
\end{figure}
\begin{figure}[htp]
		\centering
		\includegraphics[width =0.7\textwidth]{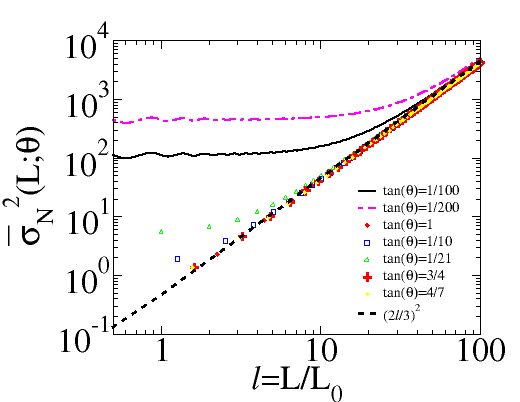}
		\caption{(color online) A log-log scale plot of $\overline{\sigma_N ^2}$ for the square lattice via square windows at rational angles.	{\color{black}All results are computed by the Monte Carlo method, except for two cases, $\tan{\theta} = 1/100,~1/200$, which are calculated from the exact expression \eqref{sigma_inte} due to the enormously large required system size.} $\fn{\overline{\sigma_N ^2}}{L;\theta}$ at rational angles are collapsed into a single scaling function, $\frac{4}{9} \left(L/L_0 \right)^2$.
		\label{fig:asymptotics_rational}}		
		\end{figure}

Now, let us examine properties of $\fn{\sigma_N ^2 }{L;\theta}$ at rational angles.
At first, $\fn{\sigma_N ^2 }{L;\theta}$ vanishes whenever side length of a square window is an integer multiple of $L_0$, i.e., $2L = n L_0 $. 
This arises because under this condition, the lengths of the square along the principal axes of the lattice are integers, and thus the number of lattice points inside the window does not change while translating the window. 
Figure \ref{fig:1/100 graph} clearly demonstrates that $\fn{\sigma_N ^2}{L;\atan{1/100}}$ vanishes at every $L = n\sqrt{10001}/2 \approx 50 n $ for an integer $n$.
Secondly, $\fn{\sigma_N ^2 }{L;\theta}$ grows like the window area, $L^2$.
More precisely, using the fact that $\fn{B}{x,y}$ is a periodic function of $x$, $\fn{\overline{\sigma_N ^2}}{L;\theta}$ is computed as
\begin{eqnarray}
\fn{\overline{\sigma_N ^2}}{L;\theta} &\approx \frac{4L^2}{9{L_0}^2} -\frac{2L}{3 {L_0}^3} +\frac{16}{45}+\frac{1}{4}\fn{H}{n,m} \label{running_asymptote} \\
& \sim \frac{4L^2}{9{L_0}^2} ~~(L\to\infty), \label{running_asymptote2}
\end{eqnarray}
where $\fn{H}{n,m}$ is the constant term of the summation in \eqref{sigma_inte}: 
	\begin{equation}\label{H}
	\fn{H}{n,m} = \frac{1}{{L_0} ^4} \sum_{k=1}^{L_0 -1} \csc{^2 \left(\pi x^{(k)}  \right)}\csc{^2 \left(\pi y^{(k)}  \right)}.
	\end{equation}
Figure \ref{fig:asymptotics_rational} depicts $\fn{\overline{\sigma_N ^2}}{L;\theta}$ computed via the Monte Carlo method at various rational angles, and they are in a good agreement with the corresponding asymptotic expression \eqref{running_asymptote}.
Note that asymptotic result \eqref{running_asymptote} is inconsistent with the spherical-window condition \eqref{HU_criterion}, which may lead one to falsely conclude that the square lattice is non-hyperuniform.
Similar phenomena also have been observed in models of two-phase heterogeneous media, e.g., the checkerboard pattern and the square lattice decorated by identical squares \cite{Anisotropy_Zachary, volumeFraction_periodic_Quintanilla}.
Specifically, even though these periodic heterogeneous media are hyperuniform by the Fourier-space condition \eqref{intro_HU_def_2phases}, the resulting volume-fraction variance decays in an anomalous fashion, i.e., $\lim_{L\to \infty} (2L)^2\fn{\sigma_V ^2}{L} \neq 0$.
Such anomalously large density fluctuations for hyperuniform systems were not predicted or noticed in previous theoretical works concerning hyperuniformity.

\begin{figure}[htp]
\begin{center}
	\begin{subfigure}[t]{0.328\textwidth}
	\includegraphics[width = \textwidth]{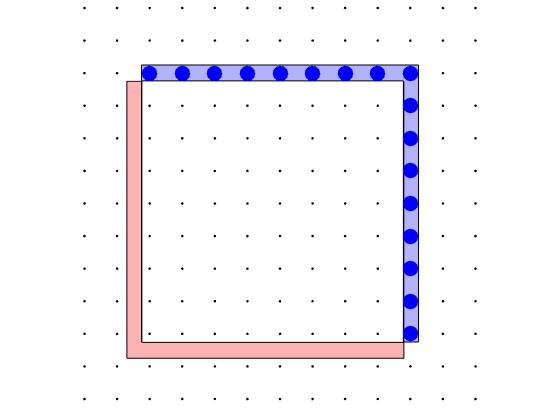}
	\caption{ }
	\end{subfigure}   
    \begin{subfigure}[t]{0.328\textwidth}
	\includegraphics[width = \textwidth]{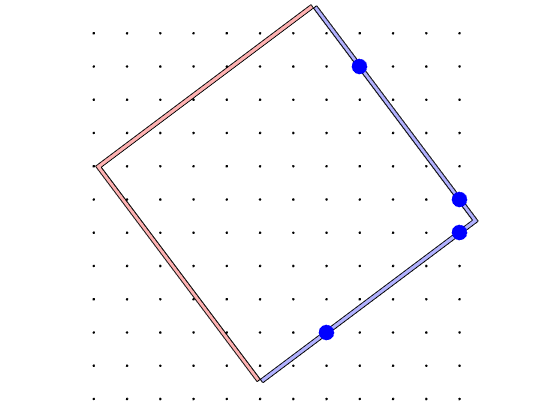}
	\caption{ }    
    \end{subfigure}
    \begin{subfigure}[t]{0.328\textwidth}
    \includegraphics[width = \textwidth]{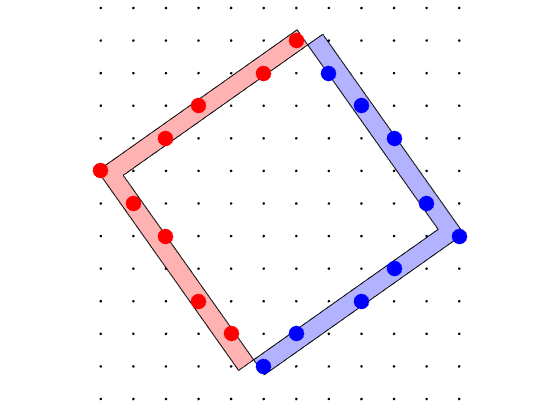}
	\caption{ }    
    \end{subfigure}
    \end{center} 
\caption{(color online) Correlation of density fluctuations for $\Z^2$ lattice points on the perimeter of the square window.
(a) $\theta = 0$ and $2L = 8.5$,
(b) $\tan{\theta} = 3/4$ and $2L = 40.5/\fn{L_0}{n,m}$, and (c) $\tan{\theta} = 1/\sqrt{2}$ and $L=7.5$. 
When the leftmost window moves toward the right upper side, the blue points fall in the window, but the red points fall out of it.
As shown in (a) and (b), at rational angles, lines of blue points can fall in the window at the same time.
However, {\color{black}in the case of irrational angles (c)}, such a correlation of the fluctuations does not happen {\color{black}because any line of lattice points cannot be parallel to a side of the window}.
\label{fig:illust_directSpace}}
\end{figure}

How do such anomalously large density fluctuations arise in what are hyperuniform systems?
We can provide two answers to this question: the first is geometrically based and the second is analytically based.
For the sake of simplicity, we will assume $\theta=0$.  
Generally speaking, anomalously large density fluctuations arise when density fluctuations on the boundary of a window are \textit{correlated}.
Specifically, for a square window, a single line of lattice points near the boundary of the window can fall alternately in and out of the window as the window moves around the lattice with a fixed orientation (see figure \ref{fig:illust_directSpace}(a) and (b)).
Thus, the resulting number variance is proportional to the square of the window perimeter in large-$L$ limit, i.e.,
\begin{equation}
\frac{\fn{\sigma_N ^2}{L;0}}{\fn{\sigma_N ^2}{a;0}} \sim  \left(\frac{L}{a}  \right)^2 ~~(L\to \infty).
\end{equation}
Roughly speaking, if the window surface (perimeter if $d=2$) has higher curvature on average or is closer to the spherical (circular) shape, then density fluctuations on the window surface are less correlated so that the growth rate of the variance becomes slower, as shown in figure \ref{fig:nv_superdisk_new}(b).
Essentially, such correlations of density fluctuations on the window surface can be demonstrated in the form of ``resonance" between $\fn{\tilde{\alpha}_2}{\vect{k};L}$ and $\fn{S}{\vect{k}}$ in the Fourier space, as shown in figure \ref{fig:FourierSpace interpretations} (a).
For the same reason, in $d$-dimensional space, the variance for the hypercubic lattice $\Z^d$ via a hypercubic window of side length $2L$ asymptotically grow like square of the window surface area in the large-$L$ limit (see \ref{sec:generalizations of aligned}):
\begin{equation}
\fn{\overline{\sigma_N ^2}}{L;0} \approx \frac{d}{6(2d-1)} (2L)^{2(d-1)}.
\end{equation}
Here, the coefficient $d$ comes from the number of faces of a $d$-dimensional hypercube.

Another way to explain anomalously large density fluctuations involves noting the conditional convergence of the second moment of total correlation function, $\int \abs{x} \fn{h}{\vect{r}} \d{\vect{r}}$ (it becomes $d$-th moment in $d$-dimensional space).  
Using the analysis in \eqref{sigma2_expansion} which was done by Torquato \etal \cite{HypU_rev}, one can asymptotically expand $\fn{\alpha_2}{\vect{r};L}$ in \eqref{sigma2_direct_square} in terms of $L$:
\begin{equation}
\fn{\sigma_N ^2}{L;0} \approx \left(2L  \right)^2 \left[\fn{A_\mathrm{square}}{L}  +\frac{\fn{B_\mathrm{square}}{L}}{L}    \right],
\end{equation}
where
\begin{eqnarray}
\fn{A_\mathrm{square}}{L} & = 1+\int_{\abs{x},\abs{y} <2L} \fn{h}{\vect{r}}\d{\vect{r}} \label{1st term_square},\\
\fn{B_\mathrm{square}}{L} & = -\int_{\abs{x},\abs{y} <2L} \abs{x}\fn{h}{\vect{r}} \d{\vect{r}}\label{second term_square}.
\end{eqnarray}
Note that the integrand $\abs{x}\fn{h}{\vect{r}}$ in \eqref{second term_square} is different from that in \eqref{sigma2_secondTerm}, $\abs{\vec{r}}\fn{h}{\vect{r}}$.
The area integral in \eqref{1st term_square} becomes an infinite sum, and its Abelian sum converges to $-1$, i.e.,
\begin{equation}\label{eq:sum_rule}
\int_{\R^2} \fn{h}{\vect{r}}\d{\vect{r}} = \lim_{\beta \to 0^+} \left[\sum_{k=1} ^\infty Z_k e^{-\beta {r_k}^2} -\frac{\pi}{\beta}	\right]=-1,
\end{equation}
where $Z_k$ stands for the coordination number of $k$-th shell of the square lattice, and $r_k$ is the radius of $k$-th shell.
Thus, $\fn{A_\mathrm{square}}{L}$ converges to 0 as $L$ tends to infinity in the sense of Abelian mean.
On the other hand, the second moment of total correlation function, given by \eqref{second term_square}, does not converge even in Abelian sum, but asymptotic behavior of its Abelian sum is $\beta^{-0.5}$ (see figure \ref{fig:convergence}).
This implies that $\fn{B_\mathrm{square}}{L} \sim L$ by the dimensional analysis of \eqref{xh(r)_convergenceTrick}.
On the other hand, the Abelian sum of the counterpart of \eqref{second term_square} for circular windows of radius $R$ converges in the large-$R$ limit \cite{HypU_rev}. More details are provided in \ref{ConvergenceTest}.

\subsection{Irrational angles}\label{sec:irrational angles}

\begin{figure}[htp]
		\centering
		\includegraphics[width = 0.75\textwidth]{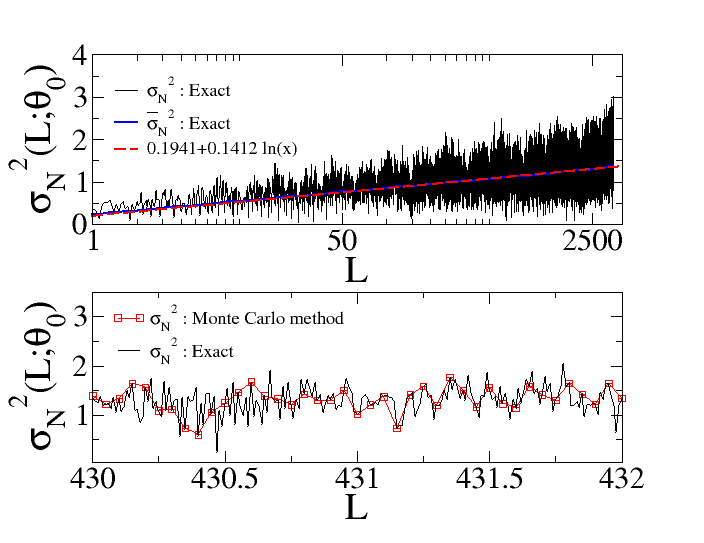}
		\caption{Semi-log plot of $\fn{\sigma_N ^2}{L;\theta_0}$ vs $L$, where $\theta_0=\tan{^{-1}\left(  1/\sqrt{2} \right)}$. Top panel shows $\sigma_N ^2$ {\color{black}obtained from the exact expression} \eqref{sigma2_direct_square}, and the cumulative moving average $\overline{\sigma_N ^2}$, which was computed by the trapezoidal rule. The red dashed line is a curve fit of {\color{black}$\fn{\overline{\sigma_N ^2}}{L;\theta_0}$ } with the natural logarithm function. The bottom panel is a magnification of a part of the top panel.
		Note that the bottom panel shows good agreement between the Monte Carlo calculations (red squares) and exact calculations (black line).\label{fig:1/sqrt(2)} }		
		\end{figure}
For a square window and the square lattice, we define an \textit{irrational angle} $\theta$ to be one that satisfies the condition $\tan{\theta} \in \R\setminus \Q$, where $\R\setminus \Q$ stands for the set of all irrational numbers.
Estimating the local number variance at irrational angles is intimately related to the concept of the Diophantine approximation in number theory and discrepancy theory in discrete mathematics \cite{Prob_Diophantine_Beck, Math_Beck_rect}.
For a given window shape (usually a rectangular box of arbitrary aspect ratios) and a finite point configuration in the unit hypercube in $\R^d$, the discrepancy is the largest difference between the window volume multiplied by the total number of points and the number of points within the window.
Generating configurations of $N$ points with the lowest discrepancy is a problem of central concern.

Currently, a general theory to analytically deal with number variances for square or rectangular windows at all irrational angles has yet to be developed.
Thus, previous works have been mainly restricted to certain types of irrational angles whose slopes have bounded partial quotients, so-called ``badly approximable numbers" (see our Supplementary Data), e.g., Fibonacci lattice \cite{Math_Dmitriy_2D_lattices}.
Beck \cite{Math_Beck_rect} studied that the variance for the square lattice with rectangular strips that have a fixed width and is tilted by an irrational angle whose slope belongs to the badly approximable numbers.
Interestingly, one can generate a large class of one-dimensional quasicrystals by projecting the square lattice points within an infinitely long rectangular strip tilted by an irrational angle onto the long axis of the strip \cite{elser_projectionMethod_quasicrystal, quasicrystal_socolar}.
Thus, the variance for the square lattice via a long rectangular strip tilted by some irrational angles is closely related to the variance for one-dimensional quasicrystals. 
\begin{figure}[htp]
    \begin{center}
    \begin{subfigure}[t]{0.48\textwidth}
    \includegraphics[width=\textwidth]{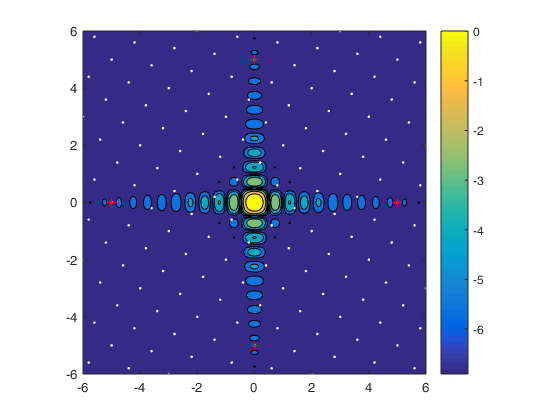}
    \caption{$L=1,~\tan{\theta} = 3/4$ \label{fig:Fourier_rational}}
    \end{subfigure}
    \begin{subfigure}[t]{0.48\textwidth}
    \includegraphics[width=\textwidth]{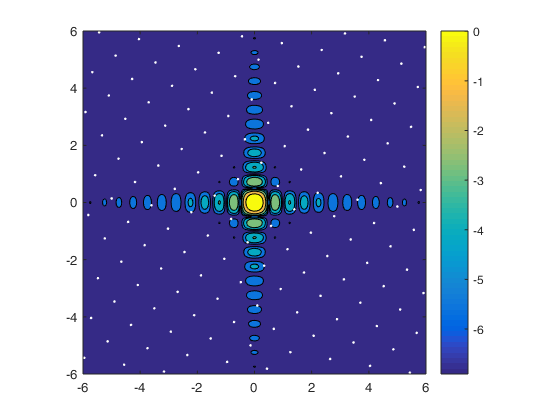}
    \caption{$L=1,~\tan{\theta} = \sqrt{2}$ \label{fig:Fourier_irrational}}
    \end{subfigure}
    \end{center} 
    \caption{(color online) Visualization of the superposition of two functions $\fn{S}{\vect{k}}$ and $\fn{\tilde{\alpha}_2}{\vect{k};L}$ that appear in the Fourier-space representation of $\fn{\sigma_N ^2}{L;\theta}$ as given by \eqref{sigma2_Fourier_square}, when $L=1$. 
    $x$ and $y$ axes of both panels are $k_x/2\pi$ and $k_y/2\pi$, respectively.
The left and right panels show the cases of a rational and irrational angles, respectively.
The white dots represent the Bragg peaks that are located on the sites of a square lattice, i.e., the structure factor $\fn{S}{\vect{k}}$.
The function $\fn{\tilde{\alpha}_2}{\vect{k};L}$, given by the formula \eqref{alpha_cubic_f}, is depicted as a contour plot in log scale for vivid visualization purposes.
The variance via \eqref{sigma2_general_Foureir} is equivalent to the summation of $\fn{\tilde{\alpha}_2}{\vect{k};L}$ at each Bragg peak.
\label{fig:FourierSpace interpretations}}
\end{figure}

We use \eqref{sigma2_direct_square} to compute $\fn{\sigma_N ^2}{L;\theta}$ at irrational angles. 
Comparing figure \ref{fig:1/100 graph} with figure \ref{fig:1/sqrt(2)},
one can see that the variance at irrational angles generally has much smaller magnitude than the variance at rational angles.
Furthermore, $\fn{\sigma_N ^2}{L;\atan{1/\sqrt{2}}}$ exhibits a logarithmic asymptotic behavior, which is similar to that for rectangular strips at certain types of irrational angles, i.e., quadratic irrational numbers \cite{Math_Beck_rect}.
This asymptotic behavior of the variance is \textit{anomalously small} in the sense that it is slower than the window perimeter, $L$.

\begin{figure}[htp]
		\centering
		\includegraphics[width = 0.75\textwidth]{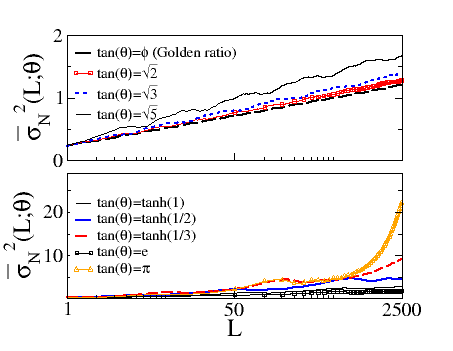}
		\caption{(color online) Cumulative moving average $\fn{\overline{\sigma_N ^2}}{L;\theta}$, vs log-scaled $L$ at various irrational angles. {\color{black}Results are calculated from the exact expression} \eqref{sigma2_direct_square}. The top panel shows the cases in which the slopes $\tan{\theta}$ are badly approximable numbers. The bottom panel shows the cases for $\tan{\theta}$ whose partial quotients are unbounded. \label{fig:sigma2_irr_comparision}}
\end{figure}

Fourier space also provides a clear way to understand the anomalously small density fluctuations for irrational angles.
Figure \ref{fig:FourierSpace interpretations} illustrates two different terms, $\fn{S}{\vect{k}}$ and $\fn{\tilde\alpha_2}{\vect{k};L}$, in the Fourier representation \eqref{sigma2_Fourier} of $\sigma_N ^2$.
For simplicity, we choose $L=1$.
If $L$ increases (not shown in the figure), the peaks of $\fn{\tilde{\alpha}_2}{\vect{k};L}$ that lie along the principal axes become narrower in their widths, and larger in their intensities.
As can be seen in figure \ref{fig:FourierSpace interpretations} (a), at rational angles, some Bragg peaks always lie along the principal axes of $\fn{\tilde{\alpha}_2}{\vect{k};L=1}$.
Then, at certain values of $L$, the peaks of $\fn{\tilde{\alpha}_2}{\vect{k};L}$ on the principal axes are coincident with those Bragg peaks, resulting in $L^2$ growth of $\sigma_N ^2$.
At irrational angles, on the other hand, there are no Bragg peaks on the principal axes of $\fn{\tilde{\alpha}_2}{\vect{k};L}$, as shown in figure \ref{fig:FourierSpace interpretations} (b).
Instead, the major contribution to the variance comes from Bragg peaks which are close to the principal axes of $\fn{\tilde{\alpha}_2}{\vect{k};L}$.
For those Bragg peaks, corresponding indices $(i,j)$ are the denominator and the numerator $(n_k,m_k)$ of convergents of $\tan{\theta}$ (see our Supplementary Data).
For this reason, it is expected that the asymptotic behavior of $\fn{\sigma_N^2}{L;\theta}$ for irrational angles largely depends on the distribution of partial quotients $a_k$, which are mainly concerned in the Diophantine approximation in number theory.
Furthermore, the variance at irrational angles is unusually small.

In terms of rational approximation, irrational numbers can be classified into two sets.
One is called \textit{``badly approximable numbers"}, including $\sqrt{n}$ and the golden ratio $\phi$.
Roughly speaking, a badly approximable number $x$ cannot have an excellent rational representation in the sense that  any rational approximation $n/m$ of $x$ approaches to $x$ at most in the order of $1/m^2$.
Thus, if $\tan{\theta}$ belongs to this set, then the Bragg peaks of the square lattice cannot be closer to the principal axes of $\fn{\tilde{\alpha}_2}{\vect{k};L}$ than a certain amount, leading to a smaller variance compared to those at irrational angles, which do not belong to badly approximable numbers.
The top panel in figure \ref{fig:sigma2_irr_comparision} shows the cumulative moving averages of the variance at some badly approximable slopes. 
Note that $\fn{\overline{\sigma_N ^2}}{L;\atan{\phi}}$ is smallest, where $\phi$ is the golden ratio, an extreme example of a badly approximable number.
The logarithmic growth rate of $\fn{\overline{\sigma_N ^2}}{L}$ was also predicted in the case of rectangular strips at angles whose slopes are quadratic irrational numbers \cite{Math_Beck_rect}, which are also ``badly approximable numbers".
The bottom panel in figure \ref{fig:sigma2_irr_comparision} shows the asymptotic behaviors of $\fn{\sigma_N ^2}{L;\theta}$ at some irrational slopes which are not badly approximable numbers.
For these angles, their asymptotic behaviors are rather unclear.

\begin{figure}[htp]
		\centering
		\includegraphics[width = 0.7\textwidth]{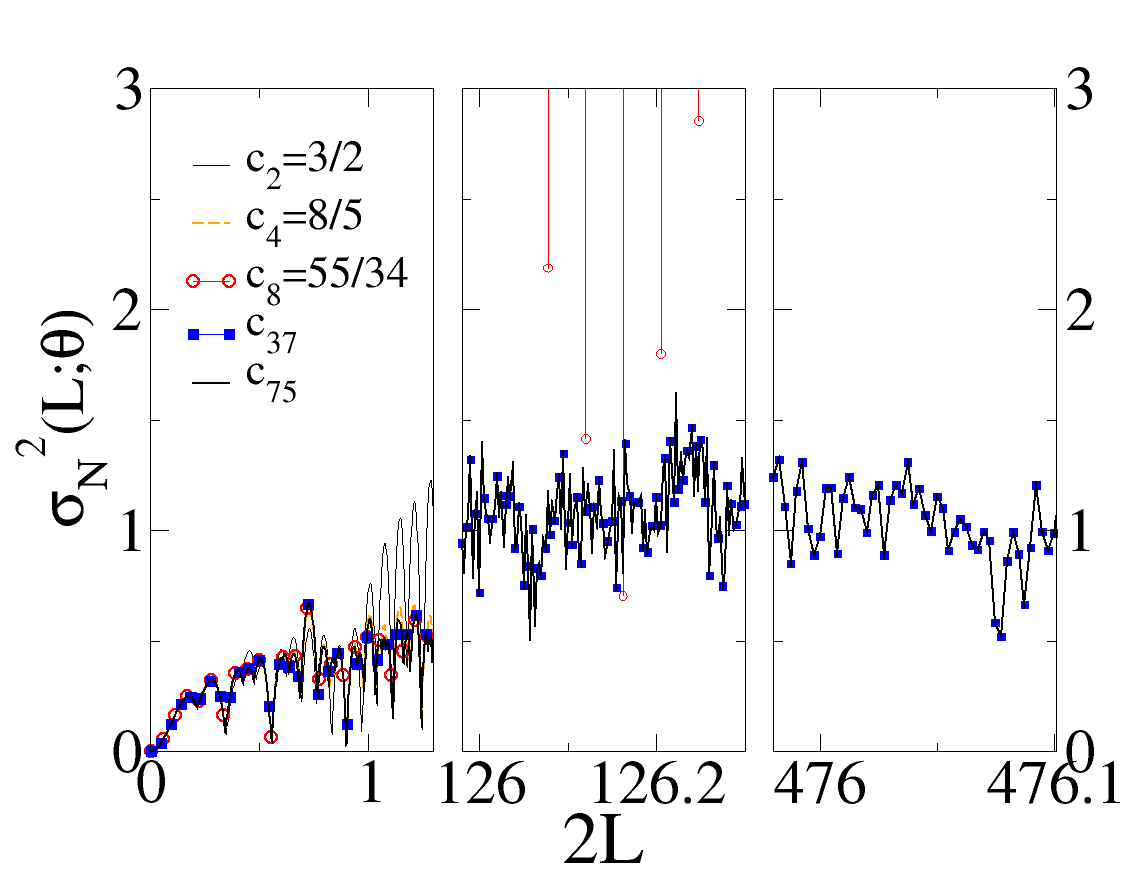}
		\caption{(color online) The variance $\fn{\sigma_N ^2}{L;\theta}$ for square windows with fixed orientations in various ranges of $L$ {\color{black}calculated by the exact expression \eqref{sigma2_direct_square}.} The angle $\theta$ between the lattice and the window is written as $\theta = \atan{c_k}$, where $c_k $ stands for $k$-th convergent of the golden ratio $\phi$, defined in our Supplementary Data. $c_{37}$ and $c_{75}$ correspond to rational approximations of $\phi$ up to the double and quadruple-precisions, respectively; see table S.A given in our Supplementary Data.		 \label{fig:rational approximation}}		
		\end{figure}
		
In addition, the difference in the asymptotic behaviors of $\fn{\sigma_N ^2}{L;\theta}$ at rational and irrational angles is one of many physical examples \cite{aubry_transition, Aubrey_transition_atomicChain, Hofstadter_Butterfly, com-incom_graphene_BN_1, com-incom_graphene_BN_2, com-incom_graphene_BN_3} at which the (in)commensurability of a certain parameter plays an critical role in physical properties.

{\bf Remarks}
\begin{enumerate}
\item There are two numerical issues in computing $\fn{\sigma_N ^2}{L;\theta}$ at irrational angles with \eqref{sigma2_direct_square}.
One is a huge round-off error. 
Relation \eqref{sigma2_direct_square} involves the subtraction of two terms on the order of $\left( 2L \right)^4$ to obtain a value on the order of unity.
Thus, a round-off error in $\fn{\sigma_N ^2}{L;\theta}$ for  $L \gtrsim 2000$ is estimated as 10\% even if we fully exploit double-precision.
For this reason, we employ quadruple-precision to compute $\fn{\sigma_N ^2}{L;\theta}$.

\item Another numerical issue is the inevitable rational approximation in an irrational angle while computing the variance at the irrational angle.
This implies that the numerically computed variances inevitably grow like $L^2$ for large $L$.
Then, it is a natural question to ask: To what extent are the numerical calculations of \eqref{sigma2_direct_square} sufficiently reliable to obtain the exact variance at irrational angles?
A qualitative answer is that for two distinct angles, the difference in variance at these angles is negligible up to a certain window size $\fn{L_\mathrm{max}}{\Delta \tan{(\theta)}}$, which is certainly a decreasing function of $\Delta \tan{(\theta)}$, as shown in figure \ref{fig:rational approximation}.
This is because \eqref{sigma2_direct_square} is a finite summation of continuous functions of $\theta$ within a range between $\fn{n_1}{i}$ and $\fn{n_2}{i}$, given by \eqref{LB of n} and \eqref{UB of n}, also determined by continuous functions of $\theta$.
\end{enumerate}

\section{Variance for disordered hyperuniform point processes}\label{sec:disordered}
Throughout this section, we solely consider a $d$-dimensional convex unit window $\omega$ of a general shape in $\R^d$ and its scalar multiplication $a \omega$, where $a$ is a positive real number.
Here, the largest distance from the centroid of $\omega$ to the boundary is the unity:
$\max_{\vect{x} \in \omega}  \abs{\vect{x}} = 1$.
The window indicator function can be written as
\begin{equation}\label{eq:window_indi_sec3}
\fn{w}{\vect{x}; \vect{R}} = \fn{w}{\vect{x}; a, \omega} = \begin{cond}
								1,& \vect{x}/a \in \omega \\
								0,&\mathrm{otherwise}.
								\end{cond}
\end{equation}
For brevity, we abbreviate the parameter set $\vect{R}$, which characterizes the window shape and orientation, to a single length-scale parameter $a$, e.g., $\fn{\sigma_N ^2}{\vect{R}}$ to $\fn{\sigma_N ^2}{a}$, $\fn{v_1}{\vect{R}}$ to $\fn{v_1}{a}$.
Then, we obtain a general expression for the asymptotic behavior of $\fn{\sigma_N ^2}{a}$ for disordered hyperuniform point processes.
In what follows, we prove that for any convex window, the variance for a disordered hyperuniform point process has the common scaling relation, which is identical to \eqref{eq:scaling relation}.
Then, we present some example calculations for two isotropic disordered hyperuniform systems.

\subsection{Analysis}\label{sec:disordered_anaylsis}
Consider statistically homogeneous and isotropic hyperuniform point processes at number density $\rho$ in $\R^d$. 
Then, the vector-dependent total correlation function becomes a radial function $\fn{h}{r} $, where $r=\abs{\vect{r}}$.
Taking advantage of the rotational symmetry, we can rewrite \eqref{sigma2_direct} for a general window shape, using the orienatationally-averaged scaled intersection volume function $\E{\fn{\alpha_2}{r; a}}_O$:
\begin{equation}\label{eq:sigma2_isotropic}
\fn{\sigma_N ^2}{a} = \rho \fn{v_1}{a} \left[1+\rho \int_{\R^d} \fn{h}{r} \E{\fn{\alpha_2}{r; a}}_O  \d{\vect{r}} \right],
\end{equation}
where $\E{\fn{\alpha_2}{r; a}}_O$ is the average of $\fn{\alpha_2}{\vect{r};a}$ over all possible orientations of $\vect{r}$ with fixed $r$.

To compute the large-$a$ asymptotic behavior of $\fn{\sigma_N ^2}{a}$, we need to find an expression for $\E{\fn{\alpha}{r;a}}_O$ for a window $a \omega$.
Generally, it is extremely difficult to find a closed expression for $\fn{\alpha_2}{\vect{r};a}$ of an arbitrarily shaped window.
For small displacements $r \ll a$, however, one can approximate it up to the first order in $\vect{r}$.
Since $\fn{\alpha_2}{\vect{r};a}$ is not differentiable at $\vect{r} = 0$, we cannot apply the multivariable Taylor theorem to $\fn{\alpha_2}{\vect{r};a}$, or equivalently $\fn{v_2 ^\mathrm{int}}{\vect{r};a}$ immediately.
Instead, we apply the Taylor theorem to $\fn{v_2 ^\mathrm{int}}{\vect{r};a}$ around $\epsilon \unitvect{r}$ and obtain a one-sided limit of the expansion as $\epsilon \to 0^+$ in the following way:
\begin{eqnarray}
 \fn{v_2 ^\mathrm{int}}{\vect{r}} & = \lim_{\epsilon \to 0^+} \left[ \fn{v_2 ^\mathrm{int}}{ \epsilon \unitvect{r}} + \sum_{\alpha=1}^d\left.\pder{\fn{v_2 ^\mathrm{int}}{\vect{r}'}}{{r'}_\alpha}\right|_{\vect{r}'=\epsilon \unitvect{r}} \left( 1 - \frac{\epsilon}{\abs{\vect{r}}}  \right)r_\alpha   + \cdots \right] \nonumber\\
 & = \fn{v_1}{a}-\fn{A_\perp}{\unitvect{r};a} \abs{\vect{r}} +\fn{O}{\abs{\vect{r}}^3}, 
\label{eq:interV_Taylor}
\end{eqnarray}
where $\fn{v_2 ^\mathrm{int}}{\vect{r}}$ is an abbreviation for $\fn{v_2 ^\mathrm{int}}{\vect{r};a}$, defined in \eqref{eq_alpha}, and $\unitvect{r}$ is the unit vector of $\vect{r}$. 
Here, the first order coefficient $\fn{A_\perp}{\unitvect{r};a}$ is defined as
\begin{equation}
\fn{A_\perp}{\unitvect{r};a} \equiv \lim_{\epsilon \to 0^+} \oint_{\partial (a \omega)} \fn{w}{\vect{x} - \epsilon\unitvect{r};a} \d{\vect{s}} \cdot \unitvect{r} ,
\end{equation}
where $\d{\vect{s}}$ represents the infinitesimal surface area element whose direction is normal to the surface and $\partial (a\omega)$ stands for the boundary of the window $a\omega$.
Geometrically, $\fn{A_\perp}{\unitvect{r};a}$ is the projected area of the window on a hyperplane normal to the displacement vector $\vect{r}$. 
The second-order Taylor coefficient of \eqref{eq:interV_Taylor} is written as
\begin{equation}\label{eq:second term _intersection volume}
\lim_{\epsilon \to 0^+} \int_{\partial (a \omega) \cap \left[\partial (a \omega) -\epsilon \unitvect{r}  \right]} \frac{\vect{x} -\epsilon\unitvect{r}}{\abs{\vect{x} -\epsilon\unitvect{r}}} \d{\vect{s}}.
\end{equation}
Note that since $\d{\vect{s}}$ in \eqref{eq:second term _intersection volume} is normal to $\vect{r}$, the second-order term in \eqref{eq:interV_Taylor} is identically zero.
Then, 
\begin{equation}\label{eq:alpha_approx_general}
\fn{\alpha_2}{\vect{r};a}\equiv \frac{\fn{v_2 ^\mathrm{int}}{\vect{r};a}}{\fn{v_1}{a}}  \approx 1 -\frac{ \fn{A_\perp}{\unitvect{r};a}}{\fn{v_1}{a}} \abs{\vect{r}}.
\end{equation}

Using the well-known average-projected-area theorem for convex bodies (see  \cite{average_projected_area_theorem}), we obtain the expression for the orientational-average of the scaled intersection volume:
\begin{equation}\label{eq:alpha_averaged}
\E{\fn{\alpha_2}{r; a}}_O \approx  1 -\frac{\E{ \fn{A_{\perp}}{\unitvect{r};a} }_O}{\fn{v_1}{a}} r =  1 -\fn{\kappa}{d} \frac{ \fn{s_1}{1}}{\fn{v_1}{1}}\frac{r}{a},
\end{equation}
where $\fn{s_1}{a}$ is the surface area of a window $a\omega$ and $\E{ \fn{A_{\perp}}{\unitvect{r};a}} _O$ stands for the orientational-average of $\fn{A_{\perp}}{\unitvect{r};a}$:
\begin{equation}
\E{ \fn{A_{\perp}}{\unitvect{r};a}} _O = \frac{1}{\Omega_d}\oint\fn{A_{\perp}}{\unitvect{r};a} \d{\unitvect{r}},
\end{equation}
where $\Omega_d = \frac{2\pi^{d/2}}{\Gamma(d/2)}$ stands for the surface area of the $d$-dimensional unit sphere.
Here, $\fn{\kappa}{d}$ is the constant depending only on spatial dimension $d$ \cite{average_projected_area_theorem}:
\begin{equation}
\fn{\kappa}{d} = \frac{\fn{\Gamma}{d/2}}{2 \fn{\Gamma}{1/2} \fn{\Gamma}{(1+d)/2}}.
\end{equation}
Using approximation \eqref{eq:alpha_averaged} and the analogous analysis in \eqref{sigma2_expansion}, we obtain
\begin{equation}\label{eq:sigma2_asymptotic_general}
\frac{\fn{\sigma_N ^2}{a}}{\fn{s_1}{a}} \approx -\rho ^2 \fn{\kappa}{d} \int_{r <2a} r \fn{h}{r}  \d{\vect{r}}~~~~(a\to \infty).
\end{equation}
Note that the right-hand side of \eqref{eq:sigma2_asymptotic_general} is independent of the window shape, thus implying that the asymptotic behavior of $\fn{\sigma_N ^2}{a}$ is independent of the window shape.

\subsection{Example calculations}\label{sec:disorder_ex}
\begin{figure}[htp]
\begin{center}
\begin{subfigure}[t]{ 0.45\textwidth}
\includegraphics[width = \textwidth]{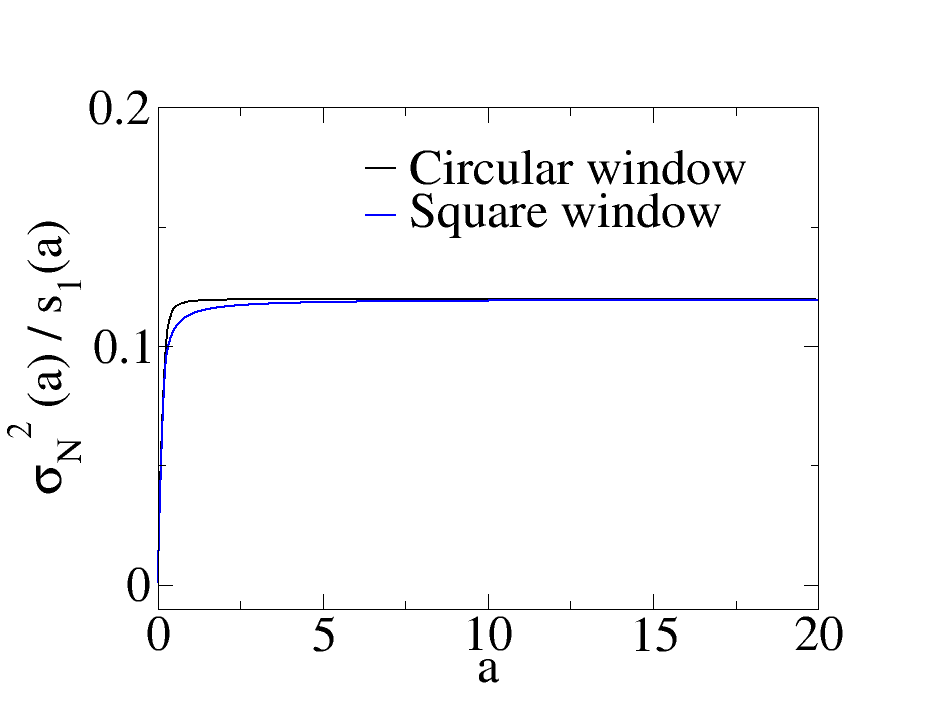}
\caption{}
\end{subfigure}
\begin{subfigure}[t]{ 0.45\textwidth}
\includegraphics[width = \textwidth]{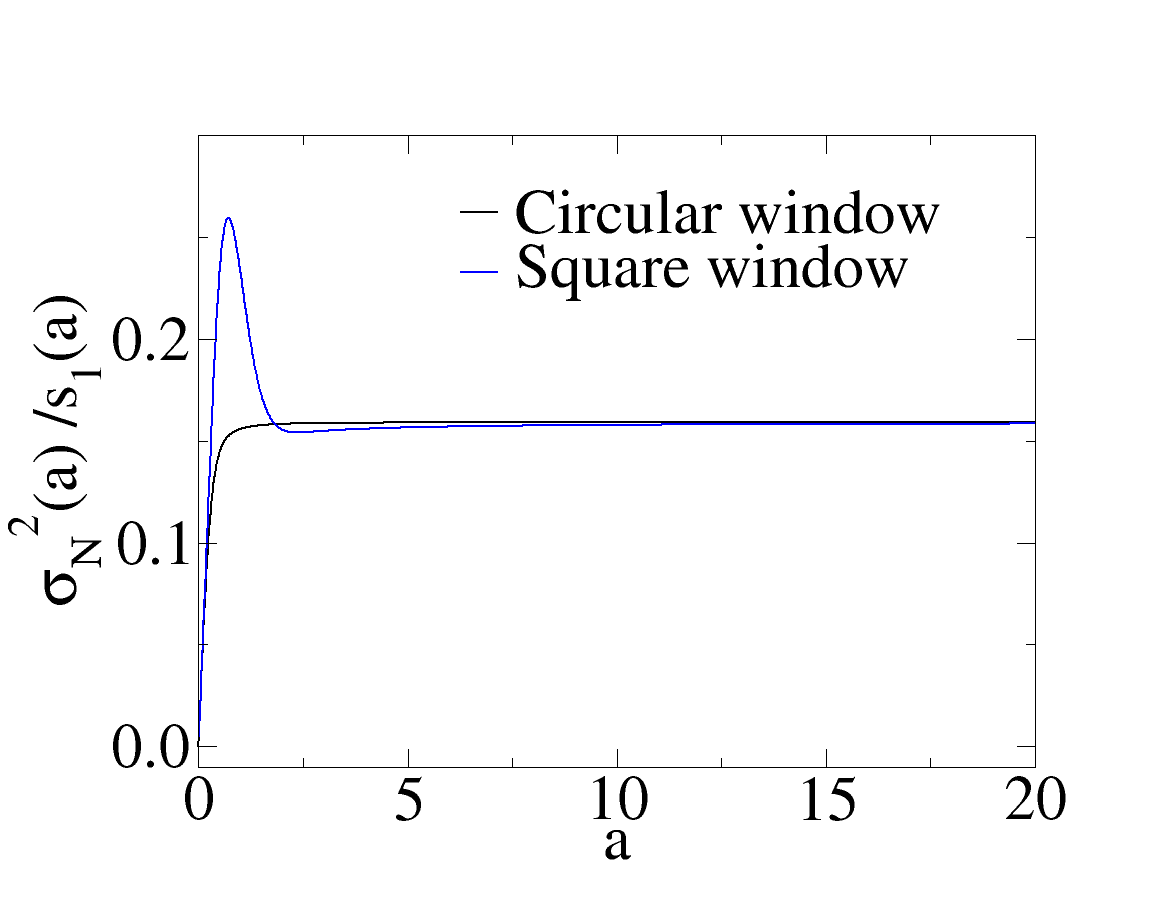}
\caption{  }
\end{subfigure}
\end{center}
\caption{ The {\color{black}exact} scaled variance $\fn{\sigma_N ^2}{a} /\fn{s_1}{a}$ for 2D disordered hyperuniform point processes for both circular and square windows as a function of $a$.
Here, $a=R$ and $a=L$ for circular and square windows, respectively, and $\fn{s_1}{a}$ denotes the window perimeter.
(a) $g_2$-step-function point process at $\rho = 1$ and $D = 1/\sqrt{\pi}$ and (b) one-component plasma at $\rho =1$ are presented.
We can see that $\fn{\sigma_N ^2}{a} / \fn{s_1}{a}$ converges to the same value as the perimeter of a window increases for each case, as shown by \eqref{eq:sigma2_asymptotic_general}.  \label{fig:DHU_asymptotics}}
\end{figure}

To demonstrate the implications of \eqref{eq:sigma2_asymptotic_general}, we study two disorderd hyperuniform point processes in $\R^2$: one is one-component plasma (OCP) \cite{HypU_rev,eg_Lebowitz_OCP} and the other is two-dimensional $g_2$-step-function point process \cite{HypU_rev}.
OCP is a system consisting of point particles of charge $e$ and uniform background charge satisfying overall charge neutrality.
In the thermodynamic limit, when the coupling constant is $\Gamma \equiv e^2 /(kT) = 2$, the total correlation function is given by \cite{exact_OCP_Jancovici}
\begin{equation}\label{eq:OCP_h}
\fn{h}{r} = -e^{-\rho \pi r^2}.
\end{equation}
Its structure factor \cite{HypU_rev} is 
\begin{equation}\label{eq:OCP_S(k)}
\fn{S}{k} \sim k^2~~ (k \to 0).
\end{equation}
We also consider a $g_2$-invariant point process defined by the following pair correlation function 
\begin{equation}\label{def:g2 step}
\fn{g_2}{r} = \fn{\Theta}{r-D},
\end{equation}
where $D$ is diameter of hard spheres.
A $g_2$-invariant process is one in which a chosen non-negative $\fn{g_2}{r}$ function remains invariant over a non-vanishing density range without changing all other macroscopic variables \cite{g2_invariantProcess1, g2_invariantProcess2}.
This system, so called $g_2$-step-function point process, turns out to be hyperuniform at the ``terminal" density $\rho = \left(\pi D^2 	\right)^{-1}$ \cite{HypU_rev}.

Since the exact expression for $\fn{g_2}{r}$ for both aforementioned point processes are given, one can compute their variance for both square and circular windows.
Figure \ref{fig:DHU_asymptotics} clearly shows that both OCP and hyperuniform $g_2$-step-function point process at the unit density have common scaling relations for circular and square windows.
It is a noteworthy fact that spherical windows measure the minimal asymptotic variance among all convex windows of the same volume.
This is because the variance is proportional to the window surface area due to \eqref{eq:sigma2_asymptotic_general} and spheres have the smallest surface-to-volume ratio among convex bodies (see Isoperimetric problems).

\section{Orientationally-averaged number variance}\label{sec:anistropic}
In the previous section, we proved that for statistically homogeneous and isotropic point processes, the asymptotic behavior of the scaled variance $\fn{\sigma_N ^2}{a}/\fn{s_1}{a}$ is independent of the window shape (see \eqref{eq:sigma2_asymptotic_general}), if windows are convex.
On the other hand, for anisotropic hyperuniform systems, including disordered ones and lattices, the growth rate of variance depends on both the shape and the orientation of windows, as we shown in \sref{sec:superdisk} and \sref{sec:lattice}.
Thus, the isotropy plays an important role in making such a difference in the asymptotic behavior.
In this section, we define orientationally-averaged local number variance and study its asymptotic behavior.

Consider statistically homogeneous but anisotropic hyperuniform point processes in $\R^d$. 
Then, we can re-interpret the variance formula \eqref{eq:sigma2_isotropic} for aspherical windows as the orientationally-averaged one.
This is because there are implicitly two different orientations in the scaled intersection volume $\fn{\alpha_2}{\vect{r};a}$ of a window $a\omega$: one is the orientation of the displacement vector $\vect{r}$ and the other is that of windows.
Thus, $\E{\fn{\alpha_2}{r; a}}_O$ also implies the average of $\fn{\alpha_2}{\vect{r};a}$ over the orientations of windows with the displacement vector $\vect{r}$ fixed.
For the purposes of illustration, consider the explicit expression \eqref{eq:visual_illustr_ang_alpha} for square windows of side length $2L$ and its scaled intersection volume $\fn{\alpha_2}{\vect{r};L}$.
The rightmost side of the first line of \eqref{eq:visual_illustr_ang_alpha} schematically represents the definition of $\E{\fn{\alpha_2}{r; a}}_O$ as the average of $\fn{\alpha_2}{\vect{r};L}$ (shaded regions) over all orientations of $\vect{r}$ (red arrows) with the orientations of windows fixed.
Rotating pairs of windows in the first line in the manner that the orientations of displacement vectors (red arrows) are identical, as shown in the second line of \eqref{eq:visual_illustr_ang_alpha}, yields the average of $\fn{\alpha_2}{\vect{r};L}$ over all orientations of square windows with a common displacement vector $\vect{r}$.
\begin{eqnarray}\label{eq:visual_illustr_ang_alpha}
\fl\E{\fn{\alpha_2}{r; a}}_O & \equiv\frac{1}{2\pi}\oint_{\abs{\vect{r}}=r} \fn{\alpha_2}{\vect{r};L}\d{\vect{r}} 
= \frac{1}{2\pi}\left[
\begin{minipage}[ht]{0.24\textwidth}
\vspace{0pt}
\includegraphics[width=\linewidth]{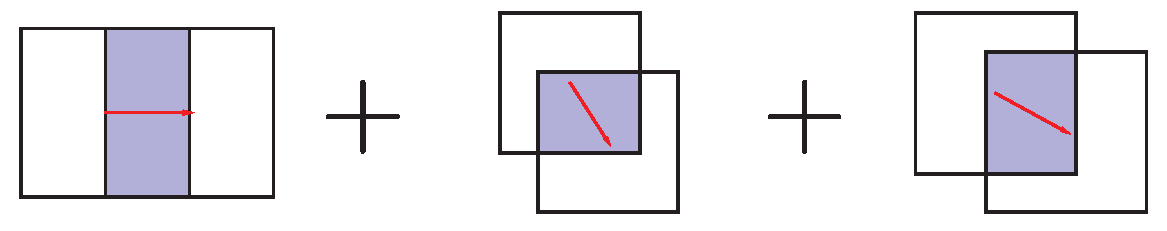}
\end{minipage}  
+\cdots 
\right] \nonumber \\
\fl& = \frac{1}{2\pi}\int_{\tiny{\begin{array}{c}\mathrm{window} \\ \mathrm{orientations} \end{array}}} \fn{ \alpha}{\vect{r}; L} \d{\theta} 
= \frac{1}{2\pi} 
\left[\begin{minipage}[ht]{0.24\textwidth}
\vspace{0pt}
\includegraphics[width=\linewidth]{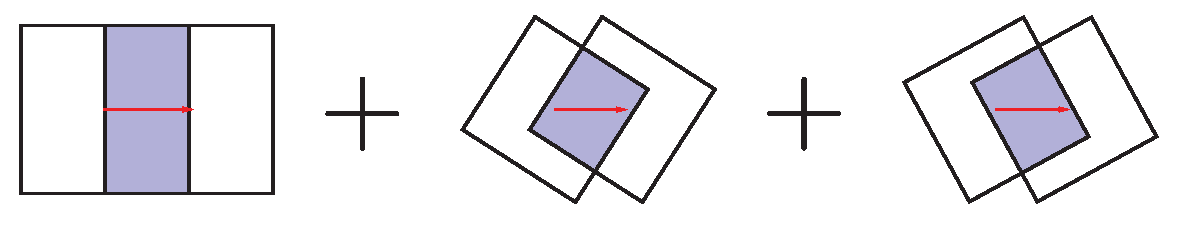}
\end{minipage}
+\cdots \right] \\
\fl&= \begin{cond}
	1-\frac{4}{\pi} x + \frac{1}{\pi}x^2, &x<1 \\
	1-\frac{2+x^2}{\pi} +\frac{4}{\pi} \left(\sqrt{x^2 -1 } -\cos{^{-1}\left( \frac{1}{x} \right)}  \right), & 1\leq x<\sqrt{2} \\
	0, &x>\sqrt{2}
	\end{cond},\nonumber
\end{eqnarray}
where $x=\frac{r}{2L}$  \cite{RHM_Torquato}.
Note that the argument made in \eqref{eq:visual_illustr_ang_alpha} is valid for any $d$-dimensional aspherical windows.
Thus, for anisotropic point processes, the expression
\begin{equation}\label{eq:ang-av sigma2}
\E{\fn{\sigma_N ^2}{a}}_O \equiv \rho \fn{v_1}{a} \left[ 1 + \int _{\R^d }\fn{h}{\vect{r}} \E{\fn{\alpha_2}{r; a}}_O \d{\vect{r}} \right]
\end{equation}
represents the orientationally-averaged variance for windows $a\omega$.
Since \eqref{eq:ang-av sigma2} is the same as \eqref{eq:sigma2_isotropic}, one can conclude that any hyperuniform point process will satisfy the following relation 
\begin{equation}\label{eq:orientation-averaged variance - asymptote}
\frac{\E{\fn{\sigma_N ^2}{a}}_O}{\fn{s_1}{a}} \approx -\rho ^2 \fn{\kappa}{d} \int_{\abs{\vect{r}} <a} \fn{h}{\vect{r}} \abs{\vect{r}} \d{\vect{r}} ~~(a \to \infty).
\end{equation}
This implies that orientationally-averaged variance $\E{\fn{\sigma_N ^2}{a}}_O$ of any hyperuniform point process will give the same scaling relation \eqref{eq:scaling relation} for any convex window shape.
Therefore, we conclude that for any convex window shape, the following hyperuniformity conditions applies:
\begin{equation}\label{HU_direct new condition}
\lim_{a \to \infty} \frac{\E{\fn{\sigma_N ^2}{a}}_O}{\fn{v_1}{a}}  = 0,
\end{equation}
which is consistent with the spherical-window condition \eqref{HU_criterion}.

\begin{figure}[htp]
		\begin{center}
		\includegraphics[width = 0.75\textwidth]{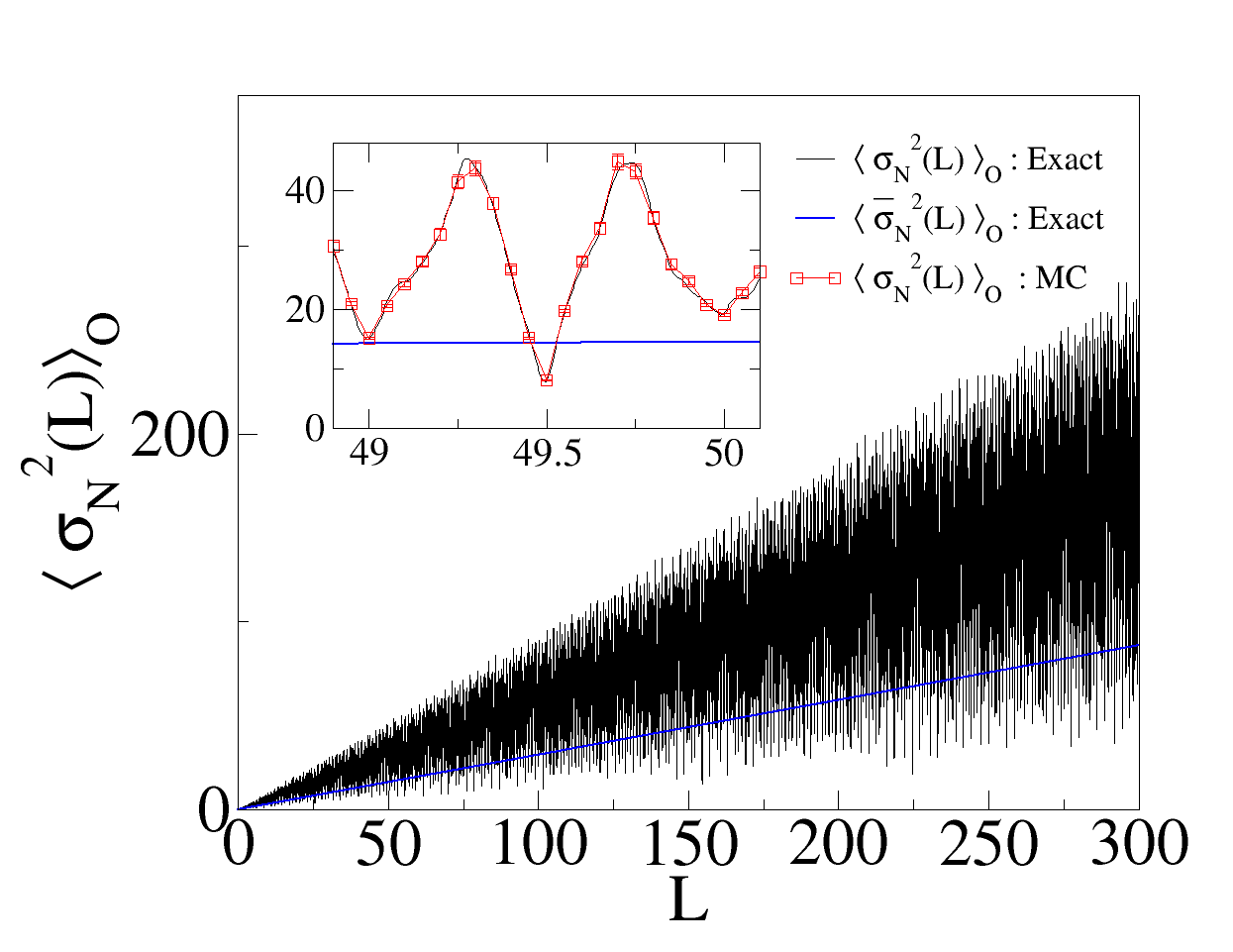}
		\end{center}
		\vspace{5pt}		
		\caption{(color online) Orientationally-averaged variance $\E{\sigma_N ^2}_O$ for the square lattice using randomly orientated square windows versus $L$. 
	Inset : comparison between exact results \eqref{eq:ang-av sigma2} and the Monte Carlo calculations of the variance $\E{\sigma_N ^2}_O $.
	In the Monte Carlo calculations, $300~000$ windows, each of which has a random centeroid and a random orientation, were thrown. \label{fig:angular_average}}		
		\end{figure}

Now, we consider the orientationally-averaged variance, $\E{\fn{\sigma_N ^2}{L}}_O$ for the square lattice using square windows.
In this case, \eqref{eq:orientation-averaged variance - asymptote} implies that
\begin{equation}\label{ang_av_asymptotic}
\lim_{L \to \infty} \frac{\overline{\E{\fn{\sigma_N ^2}{L}}_O}}{8L} = \lim_{R\to \infty} \frac{\fn{\overline{\sigma_N ^2}}{R}}{2 \pi R}.
\end{equation}
Kendall \cite{Math_Kendall_1} proved the same result for square lattice and randomly oriented planar convex windows with non-vanishing curvatures, and derived the following:
\begin{equation}\label{eq:Kendall's result}
\lim_{R\to \infty} \frac{\fn{\overline{\sigma_N ^2}}{R}}{2 \pi R} = \frac{1}{2\pi^{3}} \fn{\zeta}{\frac{3}{2}}\fn{L}{\frac{3}{2}, \chi} \approx 3.6419 \times 10^{-2},
\end{equation}
where $\fn{\zeta}{x}$ is the Riemann zeta function and $\fn{L}{x,\chi}$ is the Dirichlet $L$-function, which is defined as $\fn{L}{x,\chi} \equiv \sum_{n=0}^\infty (-1)^n/ (2n+1)^x $.
Figure \ref{fig:angular_average} clearly shows that $\overline{\E{\fn{\sigma_N ^2}{L}}_O}\sim L$, which also is consistent with Mat\'{e}rn's observation \cite{Matern1}, and from figure \ref{fig:angular_average} we obtain
\begin{equation}\label{eq:ang_av_sigma2_square_asymptotic}
\lim_{L \to \infty} \frac{\overline{\E{\fn{\sigma_N ^2}{L}}_O}}{8L} = (3.642 \pm 0.0001) \times 10^{-2}, 
\end{equation}
which is in good agreement with \eqref{eq:Kendall's result}.

This result gives a clue to the distribution of the asymptotic behaviors of $\fn{\sigma_N ^2}{L;\theta}$ with respect to the angle $\theta$.
The orientationally-averaged variance is
\begin{equation}\label{eq:ang_av_sigma2_2}
\E{\fn{\sigma_N ^2}{L}}_O =\frac{1}{\pi}\int_{-\pi/2} ^{\pi/2} \fn{\sigma_N ^2}{L;\theta}\d{\theta}.
\end{equation}
Here \eqref{eq:ang_av_sigma2_2} may not be well-defined as a Riemann integral, because for a fixed $L$, the continuity of $\fn{\sigma_N ^2}{L;\theta}$ with respect to  $\theta$ is unclear.
Therefore, it is better to introduce the probabilistic integral to compute \eqref{eq:ang_av_sigma2_2}. 
Then, since the set of rational angles, $\tan^{-1}{(\Q)}$, have zero measure among all angles, rational angles do not make any  contribution to the orientational-average of the variance, and thus
\begin{equation}\label{Lebesgue integral}
\frac{1}{\pi}\int_{-\pi/2} ^{\pi/2} \fn{\sigma_N ^2}{L;\theta}\d{\theta} = \frac{1}{\pi}\int_{[-\pi/2,\pi/2)\setminus \tan^{-1}{(\Q)} }\fn{\sigma_N ^2}{L;\theta}\d{\theta}.
\end{equation}
In order for \eqref{Lebesgue integral} to be consistent with the result \eqref{eq:ang_av_sigma2_square_asymptotic}, it is expected that $L$-growth rate for $\fn{\sigma_N ^2}{L;\theta}$ should exist for $\theta$ of non-zero measure subset of $[-\pi/2,\pi/2)\setminus \tan^{-1}{(\Q)}$.
However, we have yet to observe linear growth rate and so it is an interesting problem in mathematics to identify an irrational angle at which the variance for the square windows asymptotically grows like the window perimeter.

\section{Conclusions and discussion}\label{sec:conclusion}

We have studied the window-shape dependence on the large-window asymptotic behavior of the local number variance of hyperuniform point processes to understand conditions under which the growth rate of the variance is not slower than the window volume, conflicting with the spherical-window hyperuniformity condition \eqref{HU_criterion}.
For this purpose, we computed the variance for several hyperuniform systems using aspherical windows with a fixed orientation with respect to the systems. 

We demonstrated that for hyperuniform systems, the growth rate of the variance can depends on not only the window shape but also the window orientation. 
We begin with the numerical computation of the variance for the square lattice with superdisk windows, and demonstrate that its asymptotic behavior varies with the window shape, i.e., the deformation parameter $p$.
Importantly, as the window shape is closer to perfect squares ({\color{black}$p\to\infty$}), the asymptotic behavior of the variance approaches to $L^2$ (from $L$), which is inconsistent with ``spherical-window" condition \eqref{HU_criterion}.

Then, to better understand the conditions under which hyperuniform systems can have anomalously large variance growth in conflict with the spherical-window condition \eqref{HU_criterion}, we investigated the case of the square lattice and square windows (superdisk of {\color{black}$p\to\infty$}).
We identify two classes of angles of the square window with respect to the lattice, at which the asymptotic behavior is different.
At the rational angles, defined by \eqref{def_rational angle}, the variance for square lattice increases like the window volume.
However, at the irrational angles, the variance is significantly smaller the variance via the spherical windows.

Based on the analysis for the square window and square lattice, we explained the origin of the inconsistency in the direct-space hyperuniformity conditions for spherical and aspherical windows in two aspects.
One is the resonance between the structure factor $\fn{S}{\vect{k}}$ and the Fourier transform of the scaled intersection volume function $\fn{\tilde{\alpha}_2}{\vect{k};L}$ in the Fourier space (see figure \ref{fig:FourierSpace interpretations}).
For the square lattice and square windows, ``rational angles" are the angles at which the resonance occurs to cause the anomalously large variance growth.
Subsequently, we extended the concept of rational angles to the case of $d$-dimensional Bravais lattice and parallelepiped windows (see \ref{sec:generalizations_2d}).
We explicitly computed rational angles corresponding to square lattice and rectangular windows with a fixed aspect ratio, and the case of triangular lattice and square windows.
Another explanation is the conditional convergence of the second moment of the total correlation function, denoted by $\lim_{R\to\infty}\fn{B_N}{R}$ in \eqref{sigma2_secondTerm}.
Using Abelian summability method (\ref{ConvergenceTest}), we demonstrated that the improper integral, involved with $\lim_{R\to\infty}\fn{B_N}{R}$, is divergent for square boundaries while it is convergent for the circular one.

We proved that for statistically isotropic disordered hyperuniform systems, the variance associated with aspherical convex windows exhibits the same asymptotic behavior as the variance for spherical windows.
We verified this result for two isotropic disordered hyperuniform point processes, i.e., one-component plasma and $g_2$-step-function point process at the critical density.

We also suggest a new direct-space hyperuniformity condition that is independent of the window shape, i.e.,
\begin{equation}\label{HU_criterion_new}
\lim_{a \to \infty} \frac{\E{\fn{\sigma_N ^2}{a}}_O}{\fn{v_1}{a}}  =0 , 
\end{equation}
where $\E{\fn{\sigma_N ^2}{a}}_O$ represents the local number variance averaged over window orientations.
This is consistent with the fact that for a planar convex window, $\E{\fn{\sigma_N^2}{L}} _O$ of the square lattice is asymptotically bounded by the perimeter of the window \cite{Math_Kendall_1, Math_planar_convex_body_LatticePts_Brandolini, Math_LatticePoints_DiophantineConditions_Alex}.
We note that the same analysis and general conclusions directly extend to two-phase media because the formulas for $\fn{\sigma_N^2}{\vect{R}}$ and $\fn{\sigma_V ^2}{\vect{R}}$ are essentially the same.

We have studied how to reduce the dependence of {\color{black}the variance on the window shape} at large length scales.
For future study, it will be interesting to investigate how to design the window shape and its orientation to maximize or minimize the variance for a given system at short length scales.
Minimizing the variance corresponds to finding the ground state of the repulsive pair potential defined by the scaled window intersection volume $\fn{\alpha_2}{\vect{r};\vect{R}}$ \cite{HypU_rev}.
The results of such studies may be used in the field of self-assembly. 
For instance, in the presence of depletants, the contact attraction is exerted between two cubic nano-shells due to the osmotic pressure \cite{cubic_prt-cubic_lattice}.
Here, the attractive pair potential is proportional to the scaled intersection volume $\fn{\alpha_2}{\vect{r};L'}$, given by \eqref{alpha_cubic_d} in $d=3$, where $L'=L+R_g$, $2L$ is the side length of the cubic particle, and $R_g$ is the gyration radius of a depletant.

\section*{Note Added in Proof}
We learned recently that in the paper\cite{Martin1980}, the authors presented a formula that is the same as \eqref{eq:orientation-averaged variance - asymptote} in the present article.

\ack
This work was supported in part by the Materials Research Science and Engineering Center (MRSEC) program of the National Science Foundation under Award Number DMR-1420073.

\appendix
\section{Generalizations of rational angles to other Bravais lattices and parallelepiped windows}\label{sec:generalizations_2d}
In the case of the square lattice and square windows, we identify rational angles (\sref{sec:rational Angles}) at which the growth rate of the variance is not slower than the window volume.
The concept of rational angles (orientations in higher dimensions) can be extended to general Bravais lattices and parallelepiped windows in $d$-dimension.
For this purpose, we will derive a Fourier space representation of the variance for Bravais lattices using parallelogram observation windows in two-dimensions, and then generalize the expression to higher dimensions.	
Denote by $\fn{\Omega}{L}$ a parallelogram  window with a single length scale $L$, which is defined as
\begin{equation}
\fn{\Omega}{L} \equiv \set{\vect{x}=\sum_{i=1}^2 y_i \vect{A}_i \in \R^2}{\abs{y_i}<L,~\mathrm{for}~i=1,2},
\end{equation}
where $\vect{A}_1$ and $\vect{A}_2$ are linearly independent vectors in $\R^2$. 
The window indicator function $\fn{w}{\vect{r}; L}$ of this parallelogram window is given by
\begin{equation}
\fn{w}{\vect{r}; L} \equiv 
\begin{cond}
	1,&~\vect{r} \in \fn{\Omega}{L}\\
	0,&~\mathrm{otherwise}
\end{cond} 
= \prod_{i=1}^2 \fn{\Theta}{L-\abs{y_i}} = \fn{w_0}{M\vect{r};L},
\end{equation}
where $\fn{w_0}{\vect{r},L}$ is the window indicator function of a two-dimensional square window that has side length $2L$ and is centered at the origin, and $M$ is a linear operator that transforms the unit parallelogram $\fn{\Omega}{1}$ into the unit square.
In a matrix representation,
\begin{equation}
M = \begin{matrix_re}{c}
	{\vect{B}_1} ^T \\
	{\vect{B}_2} ^T
	\end{matrix_re},
\end{equation} 
satisfying $\vect{B}_i\cdot\vect{A}_j = \vect{B}_i ^T \vect{A}_j =\delta _{ij}$ for $i,j=1,2$, where $\delta_{ij}$ is the Kronecker delta symbol. 
The Fourier transform of the indicator function $\fn{w}{\vect{r},L}$ can be written as:
\begin{eqnarray}
	\fn{\tilde{w}}{\vect{k};L}& \equiv \int_{\R^2} \d{\vect{r}} e^{i \vect{k}\cdot\vect{r}} \fn{w}{\vect{r}; L} = \int_{\R^2} \d{\vect{r}} \exp{\left(i \vect{k}\cdot\sum_{i=1}^2 y_i \vect{A}_i   \right)}\prod_{i=1}^2 \fn{\Theta}{L-\abs{y_i}} \nonumber\\
	& =\abs{\pder{\left(r_1,r_2  \right)}{\left(y_1,y_2  \right)}} \int_{\R^2}   \d{\vect{y}} e^{i \vect{q}\cdot \vect{y}} \prod_{i=1}^2\fn{\Theta}{L-\abs{y_i}}	,\label{tilde_w step1}
\end{eqnarray}
where $\vect{r} \equiv M^{-1}\vect{y}$, and thus the Jacobian of the transformation from $\vect{r}$ to $\vect{y}$, $\abs{\pder{\left(r_1,r_2 \right)}{\left(y_1,y_2 \right)}}$ is identical to the determinant of $M^{-1}$.
Then, \eqref{tilde_w step1} becomes
\begin{equation}\label{tilde_w step2}
\fn{\tilde{w}}{\vect{k};L} = \abs{\det{(M^{-1} )}}  \fn{\tilde{w}_0}{\vect{q};L} = \abs{\det{(M^{-1} )}}  \fn{\tilde{w}_0}{\left( M^{-1} \right)^T \vect{k};L},
\end{equation}
where $\fn{\tilde{w}_0}{\vect{q};L}$ is the Fourier transform of the square window of side length $2L$, and $\vect{q} = \left(M^{-1} \right)^T \vect{k}$.
Using the convolution theorem, the Fourier transform of the scaled intersection volume function, $\fn{\tilde{\alpha}_2}{\vect{r};L}$, can be expressed as
\begin{equation}\label{general_window function}
\fl	\fn{\tilde{\alpha}_2}{\vect{k};L} =  \abs{\fn{\tilde{w}}{\vect{k};L}}^2 /\abs{\fn{\Omega}{L}} =\abs{\det{(M^{-1})}} \abs{\fn{\tilde{w}_0}{\left( M^{-1} \right)^T \vect{k} ;L}}^2 /\left( 2L \right)^2,
	\end{equation}
where $\abs{\fn{\Omega}{L}} = \left(2L  \right)^2 \abs{\det{\left( M^{-1} \right)}}$. 
In terms of reciprocal vectors $\vect{b}_i$ of lattice vectors $\vect{a}_i$, the structure factor is
\begin{equation}\label{structuraFactor2}
\fn{S}{\vect{k}}  ={(2\pi)^2}/{v_c} \sum_{\vect{n} = (n_1, n_2) \in \Z^2 \setminus \vect{0}} \fn{\delta}{\vect{k} - 2\pi B \vect{n}},
\end{equation}
where $B = \begin{matrix_re}{c c}
	\vect{b}_1~\mid~\vect{b}_2 
	\end{matrix_re}$ 
in a matrix representation, and the volume of the fundamental (unit) cell $v_c$ is equal to the inverse of the number density, i.e., $v_c = \abs{\det{\left(B^{-1}  \right)}} =1/\rho$.

Substituting \eqref{structuraFactor2} and \eqref{general_window function} into \eqref{sigma2_Fourier}, the variance can be written as
\begin{eqnarray}
\fn{\sigma_N ^2}{L;\theta} &= \frac{\rho \abs{\fn{\Omega}{L}}}{(2\pi)^2}\int_{\R^2} \d{\vect{k}} \fn{S}{\vect{k}}\fn{\tilde{\alpha}_2}{\vect{k};L}  \nonumber \\
	&	=16\frac{\abs{\det{(M^{-1})}}^2 }{{\abs{\det{(B^{-1})}}}^2}  \sum_{\vect{n}\in \Z^2 \setminus \vect{0}}\prod_{i=1}^2 \left(  \frac{\sin{ \left(2\pi \left[ \left( M^{-1} \right)^T B \vect{n} \right]_i \right)}}{2\pi\left[ \left( M^{-1} \right)^T B \vect{n} \right]_i }\right)^2  \nonumber\\
	& =  \frac{\abs{\det{(P)}}^2}{\pi^{4}} \sum_{\vect{n} \in \Z^2 \setminus \vect{0}} \prod_{i=1}^2 \left(  \frac{\sin{\left( 2\pi \left[P \vect{n} \right]_i L \right)}}{\left[P \vect{n} \right]_i} \right)^2 , \label{sigma2_generalization}
\end{eqnarray}
where $\theta$ stands for the angle between two vectors $\vect{A}_1$ and $\vect{a}_1$, $P = (M^{-1})^T B$ whose matrix element is
\begin{equation}\label{P-matrix}
 P_{ij} = \vect{A}_i ^T \vect{b}_j,
\end{equation}  
and $\left[P \vect{n}	\right]_i = {\vect{A}_i}^T \left(n_1 \vect{b}_1 +n_2 \vect{b}_2 	\right)$ for $i=1,2$.
We define $\theta$ as a ``rational angle" for any two-dimensional Bravais lattice and parallelogram windows if only one of $\left[P \vect{n}	\right]_i = 0$ has a non-trivial integral solution $\vect{n}$.
Note that $\fn{\sigma_N ^2}{L;\theta}$ grows like $L^2$ at such an angle $\theta$, to be contrasted with the spherical-window condition \eqref{HU_criterion}.
Straightforwardly, one generalizes \eqref{sigma2_generalization} to $d$-dimension as
\begin{equation}\label{sigma2_generalization_d}
\fn{\sigma_N ^2}{L; P} = \frac{\abs{\det{(P)}}^2}{\pi^{2d}} \sum_{\vect{n} \in \Z^d \setminus \vect{0}} \prod_{i=1}^d \left(  \frac{\sin{\left( 2\pi \left[P \vect{n} \right]_i L \right)}}{\left[P \vect{n} \right]_i} \right)^2,
\end{equation}
and the window orientations with respect to the lattice is characterized by the $d \times d$ matrix $P$, defined in \eqref{P-matrix}.
We say that that the window is at a rational orientation if $P$ has a non-trivial integral solution $\vect{n} \in \Z^d\setminus{\vect{0}}$ satisfying that a vector $P\vect{n}$ has at least $d/2$ vanishing elements.

{\bf Remarks}
\begin{enumerate}
\item For the square lattice and for the square windows is rotated by $\theta$ with respect to the lattice, $M$ is the identity matrix, and $B=R^T$, where $R$ is given by \eqref{RotationMatrix}. 
Then, we can recover \eqref{sigma2_general_Foureir}.

\item If both window and lattice are spanned by the same basis vectors and are aligned, i.e., $\theta = 0$ and $P_{ij} =\delta_{ij}$,
then \eqref{sigma2_generalization} becomes identical to $\fn{\sigma_N ^2}{L;\theta}$ of the square lattice using a square window of side length $2L$, given by \eqref{sigma2_0}, up to a proportional constant. 
However, if the lattice is rotated with respect to the window, then operator $P$,  defined in \eqref{P-matrix}, can be written as
\begin{equation}
P = \left( M^T \right)^{-1}R M^T,
\end{equation}
where $R$ is a rotation operator. 
Thus, $P$ is a similarity transform of the rotation operator $R$ in this case.

\end{enumerate}

\subsection{Square lattice and rectangular windows }\label{app:square lattice and rectangular window}
The generalization to a rectangular window and the square lattice is straightforward. 
For a rectangular window, rotated by an angle $\theta$, of two side lengths $aL$ and $bL$,
\begin{equation}
P = \begin{matrix_re}{cc}
	a & 0\\
	0 & b
	\end{matrix_re} R,
\end{equation}
and thus $\left[P\vect{n}  \right]_1 = a \left[ R\vect{n} \right]_1$ and $\left[P\vect{n}  \right]_2 = b \left[ R\vect{n} \right]_2$, where $R$ is the rotation matrix, defined by \eqref{RotationMatrix}.
Therefore, \eqref{sigma2_generalization} becomes
\begin{equation}\label{sigma2_rectangle}
	\fn{\sigma_N ^2}{L;\theta} =  \frac{1}{\pi^{4}} \sum_{\vect{n} \in \Z^2 \setminus \vect{0}} \left(  \frac{\sin{\left( 2\pi \left[R \vect{n} \right]_1 a L \right)}}{\left[R \vect{n} \right]_1} \right)^2\left(  \frac{\sin{\left( 2\pi \left[R \vect{n} \right]_2 bL \right)}}{\left[R \vect{n} \right]_2} \right)^2 .
\end{equation}
Note that \eqref{sigma2_rectangle} is similar to its counterpart to square windows \eqref{sigma2_Fourier_square}.
Thus, one can immediately notice that for the square lattice, rational angles for both rectangular and square windows are the same, and the variance at the rational angle is
\begin{eqnarray}
\fn{\sigma_N ^2}{L;\theta}& = \frac{\left( 2ax \right)^2}{{L_0}^4}\fn{g}{2bx}+\frac{\left( 2bx \right)^2}{{L_0}^4}\fn{g}{2ax} +\frac{\fn{g}{2ax}\fn{g}{2bx}}{{L_0}^4} \nonumber \\
 &+\frac{1}{{L_0} ^4}\sum_{k=1}^{{L_0}^2 -1} \fn{B}{ax,x^{(k)}}\fn{B}{bx,y^{(k)}}, \label{sigma_rectangular window at rational angles} \label{sigma2_rectangle_rational angle}
\end{eqnarray}
where we use the same notations that we defined in \eqref{sigma_inte}.
We note that \eqref{sigma_rectangular window at rational angles} asymptotically increases like $L^2$.
The same result was derived by Rosen \cite{Math_Rosen_rationalRectangle} using a different approach.

 \subsection{Triangular lattice and square windows}\label{app:triangular lattice and square window}
Consider a triangular lattice of lattice constant $a$ with whose number density $\rho=\frac{2}{\sqrt{3}a^2}$ and square windows of side length $2L$.
Let the lattice vectors be specified by
\begin{equation}
\vect{a}_1 = a\begin{matrix_re}{cc}
				1 & 0 
				\end{matrix_re}^T	
,~~~~~~~~~\vect{a}_2 = a\begin{matrix_re}{cc}
				1/2 & \sqrt{3}/2 
				\end{matrix_re}^T	 ,
\end{equation}
thus the corresponding reciprocal vectors are given by
\begin{equation}
\vect{b}_1 = a^{-1}\begin{matrix_re}{cc}
						1 & -1/\sqrt{3}
						\end{matrix_re}^T ,
~~~~~~\vect{b}_2 = a^{-1}\begin{matrix_re}{cc}
						0 & 2/\sqrt{3}
						\end{matrix_re}^T	 .					
\end{equation}
For simplicity, consider that the principal axes of square windows are aligned along axes of Cartesian coordinates, and the lattice is rotated counterclockwise by $-\theta$ with respect to the square window. 
Then, the matrix $P$, defined by \eqref{P-matrix}, will be 
\begin{equation}
P = IB = R^T \begin{matrix_re}{cc}
		\vect{b}_1 & \vect{b}_2
		\end{matrix_re}
		=\frac{1}{a}\begin{matrix_re}{cc}
		\cos{\theta}-\frac{1}{\sqrt{3}}\sin{\theta} & \frac{2}{\sqrt{3}} \sin{\theta} \\
		-\sin{\theta}-\frac{1}{\sqrt{3}}\cos{\theta} & \frac{2}{\sqrt{3}} \cos{\theta} 
\end{matrix_re}	.
\end{equation}

Here, we can obtain two types of ``rational angles" in this case:
\begin{equation}\label{triangle_condition}
\tan{\theta} = \sqrt{3}n/(n-2m), ~(2m-n)/(\sqrt{3}n) 
\end{equation}
for integers $n$ and $m$. 
For a given rational angle $\theta$, defined by \eqref{triangle_condition}, there are at most two pairs of coprime integers $\vect{n}_i =(n_i, m_i)$ for $i=1,2$: 
\begin{equation}\label{triangle_pair}
m_1 /n_1=\left(1+\sqrt{3}\tan{\theta} 	\right)/2, ~~m_2/n_2 = \left( 1-\sqrt{3} \cot{\theta}	\right)/2.
\end{equation}
Then, the relation between $\vect{n}_1$ and $\vect{n}_2$ can be obtained:
\begin{equation}\label{secondPair}
m_2 /n_2   = \left( m_1 -2n_1	\right)/ \left( 	2m_1 -n_1\right).
\end{equation}
Thus, we can define two length scales $L_1,~L_2$ at a given rational angle $\theta$:
\begin{equation}
L_i \equiv \abs{\left[P \vect{n}_i  \right]_i } = \frac{2}{\sqrt{3} a} \sqrt{n_i ^2 -n_i m_i +m_i ^2},
\end{equation}
for $i=1,2$, and these two length scales are related in the following way:
\begin{equation}\label{L1-L2 relation}
L_2 = \sqrt{3} \abs{\frac{n_2}{2m_1 - n_1}}L_1 = \sqrt{3}L_1.
\end{equation}
If neither $m_2$ and $n_2$ is zero, $\abs{m_1 -2n_1}$ and $\abs{2m_1 -n_1}$ in \eqref{triangle_pair} are coprime due to the working principle of Euclidean algorithm (see \cite{Math_Stark_NumberTheory}).
Because of the uniqueness of the irreducible fraction, $\abs{n_2} = \abs{2m_1 -n_1}$ and $\abs{m_2} = \abs{m_1 -2n_1}$, which leads the last equality in \eqref{L1-L2 relation} to be valid.
\Tref{tab:ratinoalAngles_triLatt_sqWin} lists some $\vect{n}_1$ and $\vect{n}_2$ pairs.
\begin{table}[ht!]
\centering
\begin{tabular}{c|c c c c|c c}
\hline 
$\tan{\theta}$&$m_1$ & $n_1$ & $m_2$ & $n_2$ & $L_1$ & $L_2$ \\ 
\hline 
$1/\left(3\sqrt{3} 	\right)$&2 & 3 & -4 & 1 & $2\sqrt{7/3}$ & $2\sqrt{7}$ \\ 
\hline 
$1/\left( 7\sqrt{3}	\right) $	&	4 & 7 & -10 & 1 & $2\sqrt{37/3}$ & $2\sqrt{37}$ \\ 
\hline 
$11/\left( 9\sqrt{3}\right) $ & 10 & 9 & -11 & 8 & $2\sqrt{91/3}$ & $2\sqrt{91}$ \\ 
\hline 
$13/\left( 7\sqrt{3}\right)$ & 10 & 7 & -4 & 13 & $2\sqrt{79/3}$ & $2\sqrt{79}$ \\
\hline
$17/\left(3\sqrt{3}\right)$ & 10 & 3 & 4 & 17 & $2\sqrt{79/3}$ &$2\sqrt{79}$ \\
\hline
\end{tabular}
\caption{$\vect{n}_1$ and $\vect{n}_2$ pairs for rational angles in the triangular lattice and the square windows. \label{tab:ratinoalAngles_triLatt_sqWin}}
\end{table}

\begin{figure}[htp]
\begin{center}
\begin{subfigure}[t]{  0.49\textwidth}
\includegraphics[width = \textwidth]{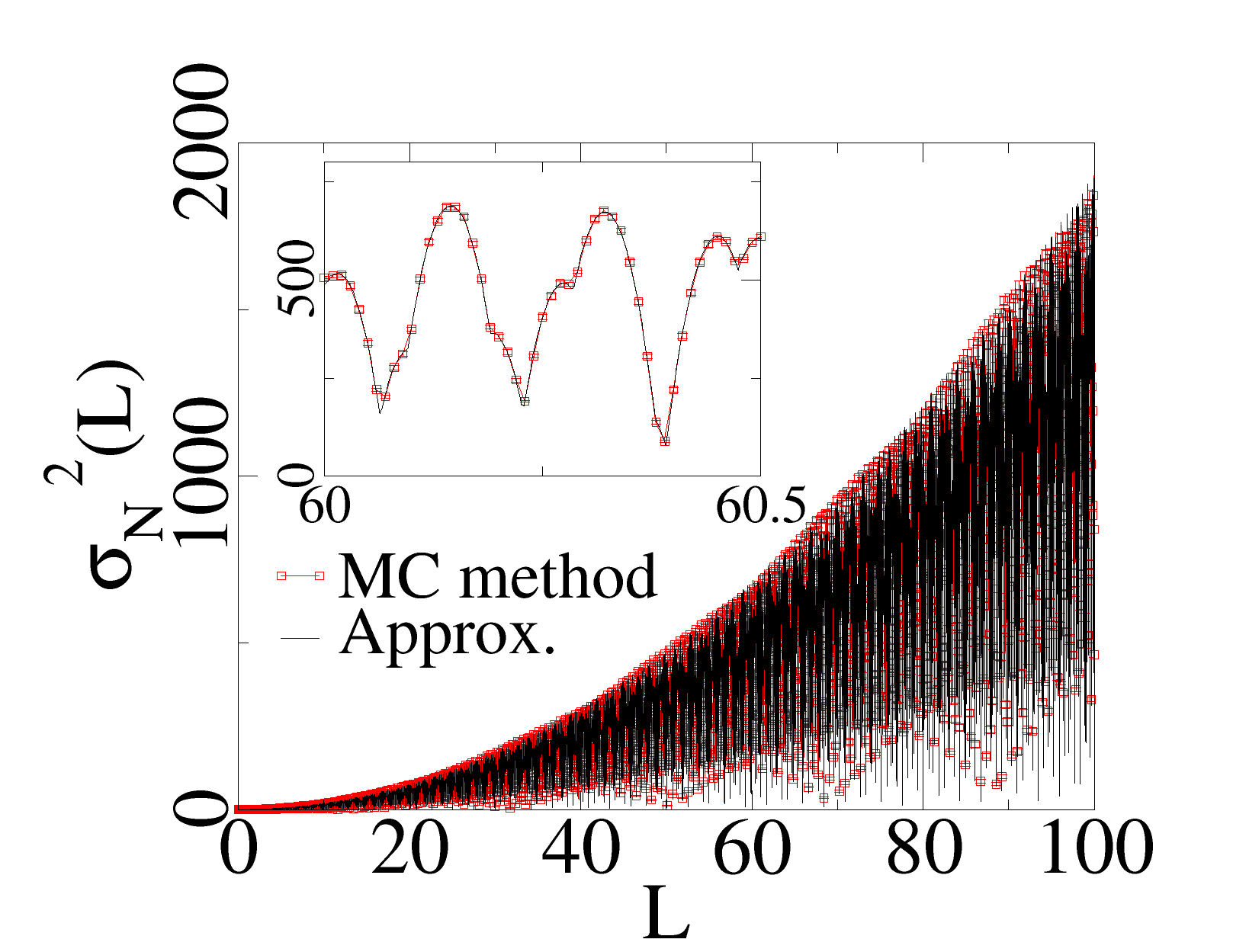}
\caption{   }
\end{subfigure}
\begin{subfigure}[t]{  0.49\textwidth}
\includegraphics[width = \textwidth]{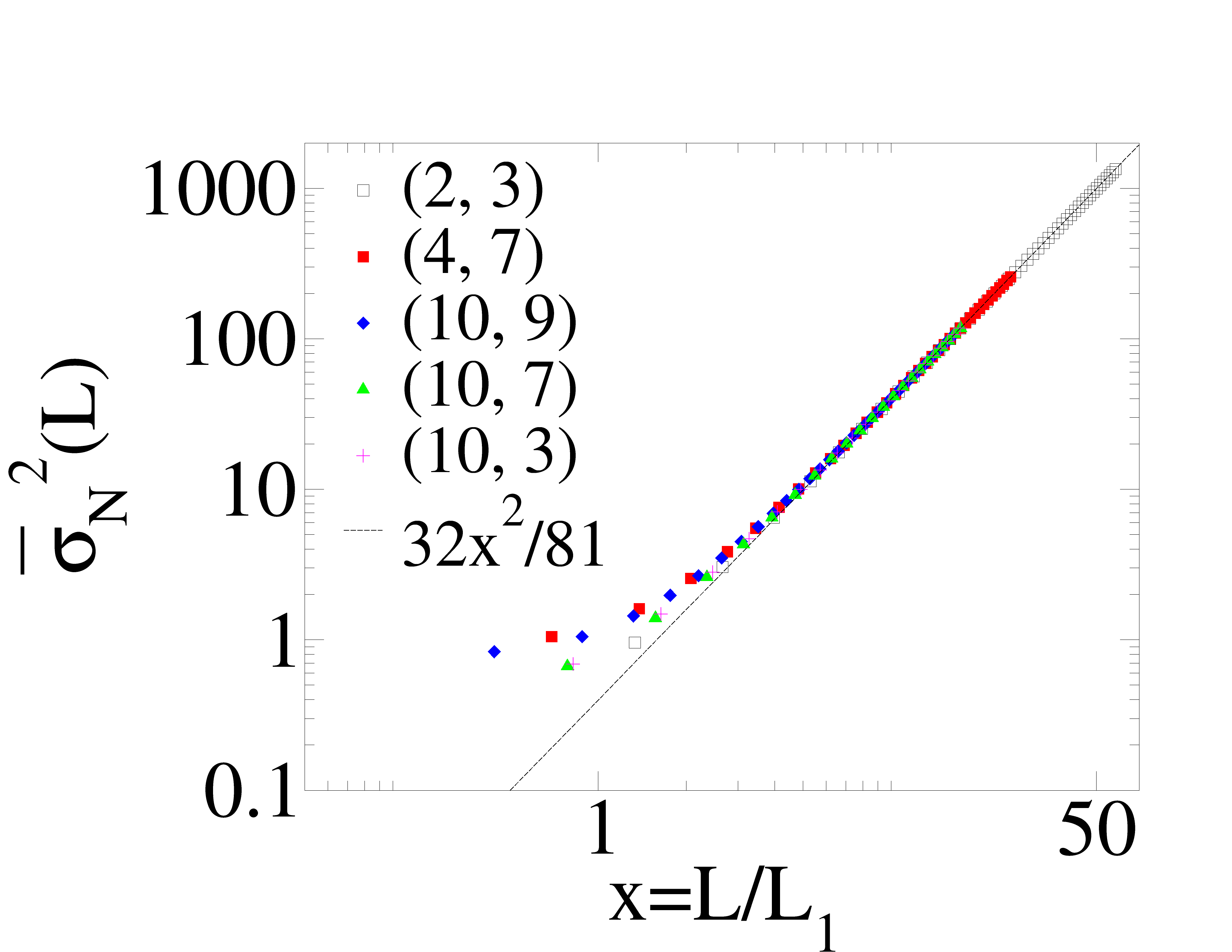}
\caption{    }
\end{subfigure}
\end{center}
\caption{The variance for the triangular lattice of the unit lattice constant using square windows.
(a) The case of $\tan{\theta} = 1/\sqrt{27}$, i.e, $(m_1,n_1)=(2,3)$. This clearly demonstrates good agreement between the approximation \eqref{sigma2_triangle} and the Monte Carlo calculations.
(b) A log-log plot of cumulative moving average $\overline{\sigma_N ^2}$ vs $L/L_1$ for various window orientations.
{\color{black}Values are computed by the Monte Carlo method using $40~000$ sampling windows with a fixed orientation.}
Here, the integer pairs in the legend represent $(m_1, n_1)$ in table \ref{tab:ratinoalAngles_triLatt_sqWin}.
These values collapse onto a single scaling function $32x^2 /81$, i.e., \eqref{sigma2_runn+triangle} in the large-$L$ limit..
\label{fig:triangular_lattice}}
\end{figure}

Then, we can find an approximation of \eqref{sigma2_generalization} at rational angles: 
\begin{eqnarray}
\fl\fn{\sigma_N ^2}{L;\theta} &\approx \frac{ \abs{\det{(P)}}^2}{ \pi^{4}} \left(  2\pi L\right)^2\sum_{k=1}^\infty \left[ 2\frac{\sin^2{\left(2\pi \left[P \left(k \vect{n}_1  \right)  \right]_1 L  \right)}}{ {\left[P \left(k \vect{n}_1  \right)  \right]_1 }^2}+2\frac{\sin^2{\left(2\pi \left[P \left(k \vect{n}_2  \right)  \right]_2 L  \right)}}{{\left[P \left(k \vect{n}_2  \right)  \right]_2 }^2} \right] \nonumber \\
\fl& = \left(\frac{2}{\sqrt{3}a^2}  \right)^2 \left( 2L \right)^2 \left[ \frac{\fn{g}{2L_1 L}}{{L_1} ^2} +\frac{\fn{g}{2L_2 L}}{{L_2} ^2}  \right] ~~(L \to \infty), \label{sigma2_triangle}
\end{eqnarray} 
where an identity, $\fn{g}{x} \equiv \fra{x}(1-\fra{x}) =2 \sum_{n=1}^\infty \left( \sin{\left( n\pi  x \right)} /(n\pi)	\right)^2$, is used.
Figure \ref{fig:triangular_lattice}(a) shows that good agreement between the approximation \eqref{sigma2_triangle} and the Monte Carlo calculations.
The asymptotic behavior of $\fn{\overline{\sigma_N ^2}}{L;\theta}$ of the approximation is given by \eqref{sigma2_triangle}:
\begin{equation}\label{sigma2_runn+triangle}
\fn{\overline{\sigma_N ^2}}{L} \approx \left(\frac{2}{\sqrt{3}a^2}  \right)^2 \frac{2}{9}\left(  \frac{1}{{L_1} ^2} +\frac{1}{{L_2} ^2}\right)L^2 = \frac{32}{81} \left(\frac{L}{aL_1}  \right)^2 ~~(L\to\infty).
\end{equation}
Figure \ref{fig:triangular_lattice}(b) illustrates that the cumulative moving averages at five different angles collapse onto a single scaling function, as we saw in figure \ref{fig:asymptotics_rational}.

\section{Generalizations of the expresion \eqref{sigma2_0} to the higher dimensions}\label{sec:generalizations of aligned}
\begin{figure}[htp]
		\centering
		\includegraphics[width =0.6\textwidth]{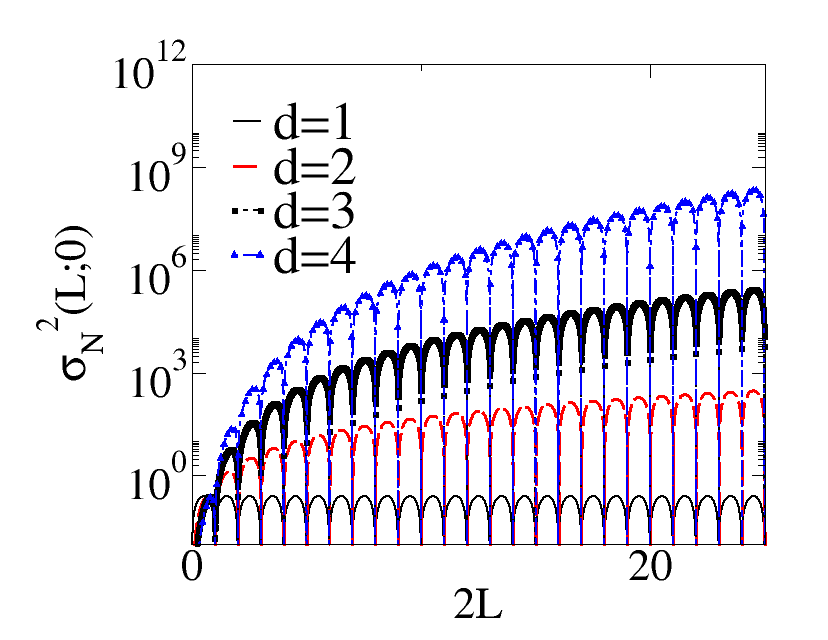}
		\caption{A semi-log plot of $\fn{\sigma_N ^2}{L;0}$ vs $2L$ {\color{black}for the first 4 dimensions. Graphs are obtained from the exact expression \eqref{sigma2_0_d dimensions}. Note that $\fn{\sigma_N ^2}{L;0}$ vanishes whenever $2L$ is an integer.} \label{fig:nv_higher dimensions}}		
\end{figure}

The generalizations of the variance for the square lattice using perfectly-aligned square windows, \eqref{sigma2_0}, to the higher dimensions are straightforward.
In $d$-dimension, \eqref{alpha_cubic-d} becomes
\begin{equation}\label{alpha_cubic_d}
\fn{\alpha_2}{\vect{r};L} \equiv \frac{\fn{v_2 ^\mathrm{int}}{\vect{r};L}}{(2L)^d} =\prod_{i=1}^d \left( 1-\frac{\abs{r_i}}{2L} 	\right) \fn{\Theta}{2L-\abs{r_i}}.
\end{equation}
Using \eqref{alpha_cubic_d}, the variance \eqref{sigma2_direct_square} can be readily generalized to the $d$-dimensional case:
\begin{eqnarray}
\fn{\sigma_N ^2}{ L;0} &= (2L)^d \left[ 1-(2L)^d  - \fn{\alpha_2}{\vect{0}; L} +\sum_{\vect{n}\in \Z^d} \fn{\alpha_2}{\vect{n};L}  \right]     \nonumber \\
	&=(2L)^d \left[-(2L)^d+\left(\sum_{n=-\floor{2L}}^{\floor{2L}} \left( 1-\abs{n}/(2L) \right) \right)^d  \right] \nonumber\\
	&= \left((2L)^2+\fn{g}{2L}  \right)^d -(2L)^{2d} .\label{sigma2_0_d dimensions}
\end{eqnarray}
Its asymptotic expression, $\fn{\overline{\sigma_N ^2}}{L}$, is
\begin{equation}\label{running-d dimension}
\fn{\overline{\sigma_N ^2}}{L;0} \approx \frac{d}{6(2d-1)} (2L)^{2(d-1)}~~(L\to\infty).
\end{equation}
Note that the variance increases like the square of the window surface area, and the coefficient $d$ in \eqref{running-d dimension} can be interpreted as the maximum number of faces in which density fluctuations can occur while the window moves.
Figure \ref{fig:nv_higher dimensions} shows $\fn{\sigma_N ^2}{L;0}$ for hypercubic lattice for the first four dimensions.

\section{Convergence of the second moment of total correlation function in two-dimensional space}\label{ConvergenceTest}
There are several methods of summability methods to assign a finite value to an infinite sequence that is not convergent in the conventional sense.
Here, we will briefly introduce two summability methods, Ces\`{a}ro and Abelian means, to explain the anomalously large density fluctuations of $\fn{\sigma_N ^2}{L;0}$ in \eqref{sigma2_0}.

\subsection{Ces\`{a}ro summability}
For a given real-valued function $\fn{f}{x}$ defined in $\R$, its improper integral $\int_0 ^\infty \fn{f}{x}\d{x}$ is called Ces\`{a}ro summable or, equivalently, $(C,\alpha)$ summable to a real number $I$, if
\begin{equation}\label{def_Cesaro sum}
\lim_{t\to \infty} \int_0 ^t \left( 1-\frac{x}{t}	\right)^\alpha \fn{f}{x} \d{x} = I,
\end{equation}
$\alpha > 0$. 
$(C,0)$ summable is equivalent to the conventional convergence of the given improper integral $\int_0 ^\infty \fn{f}{x}\d{x}$. 
$(C,1)$ summability of the improper integral is the same as the convergence of the cumulative moving average of the integral \cite{Math_summability}:
\begin{equation}\label{C,1 convergence}
\lim_{t\to \infty} \int_0 ^t \left( 1-\frac{x}{t}	\right) \fn{f}{x} \d{x} = \lim_{t \to \infty} \frac{1}{t} \int_0 ^t \left( \int_0 ^{x_2} \fn{f}{x_1} \d{x_1}	\right)\d{x_2}.
\end{equation}
Using mathematical induction, one can easily show that for a natural number $n$, an improper integral $\lim_{t\to\infty}\int_0 ^t \fn{f}{x}\d{x}$ is $(C,n)$ summable to $I$, if and only if its $n$-th multiple cumulative moving average converges to $I/{n!}$ in the large-$t$ limit:
\begin{equation}
\lim_{t\to\infty} \int_0 ^t \left(1-\frac{x}{t} 	\right)^n \fn{f}{x}\d{x} = \lim_{t\to \infty}\frac{n!}{t^n} \int_0 ^t \d{x_n} \int_0 ^{x_n} \cdots \int_0 ^{x_1} \d{x}\fn{f}{x}.
\end{equation}
Note that for spherical windows of radius $R$, $\fn{\sigma_N ^2}{R}$ for the square lattice is asymptotically linear in $R$ in the large-$R$ limit in the sense that \cite{Math_Kendall_1}
\begin{equation}\label{integral}
\lim_{R \to \infty} \frac{1}{R} \int_1 ^R \frac{\fn{\sigma_N ^2}{r}}{r} \d{r} 
\end{equation} 
converges to the constant.
Using the analysis in \eqref{sigma2_expansion}, one can express \eqref{integral} as
\begin{equation}
\sim \lim_{R\to\infty} \frac{-1}{2\fn{\Gamma}{1/2}\fn{\Gamma}{3/2}R} \int_1 ^R  \d{r}\int_0 ^r \d{r'} {r'}^2 \fn{h}{r'},
\end{equation}
which implies that the $2$nd moment of the total correlation function of the square lattice is $(C,1)$ summable in the circular boundary.

\subsection{Abelian summation}\label{sec:abelian sum}
Suppose that $\lambda = \left\{\lambda_0, \lambda_1, \lambda_2, \cdots	\right\}$ is a strictly increasing sequence approaching infinity, and $\lambda_0 > 0$. 
The Abelian mean $A_\lambda$ of a sequence $c =\left\{c_0, c_1, c_2 ,\cdots	\right\}$ is defined as
\begin{equation}
A_\lambda = \lim_{\beta \to 0^+} \fn{f}{\beta},
\end{equation}
where $
\fn{f}{\beta} \equiv \sum_{n=0}^\infty c_n e^{-\beta \lambda_n }$, and $\fn{f}{\beta}$ is assumed to be convergent for all real numbers $\beta >0$.
Abelian summation of a sequence $a$ is a special case of its Abelian mean in which $\lambda_n = n$, and thus $
\fn{f}{\beta} = \sum_{n=0 }^\infty  c_n e^{-\beta n}$.

\subsection{Conditional convergence of the integral involving the total correlation function.}
For the square lattice, the integrals of total correlation function, \eqref{sigma2_1stTerm} and \eqref{1st term_square}, oscillate but do not converge in the conventional sense.
These integrals are summable via an Abelian sum, which turns out to be equivalent to the ``convergence trick" \cite{HypU_rev}. 
For both square and circular windows, the volume integral $\int \fn{h}{\vect{r}}\d{\vect{r}}$ can be written in the same form (see \eqref{eq:sum_rule})
\begin{equation}\label{eq:Abelian_sum rule}
\fl\int_{\R^2} \fn{h}{\vect{r}} \d{\vect{r}}= \lim_{\beta \to 0^+} \int_{\R^2} \fn{h}{\vect{r}}e^{-\beta r^2} \d{\vect{r}} = \lim_{\beta \to 0^+} \left[\sum_{k=1}^\infty Z_k e^{-\beta {r_k}^2}-\frac{\pi}{\beta}  \right] = -1.
\end{equation}

\begin{figure}[htp]
		\centering
		\includegraphics[width =0.6 \textwidth]{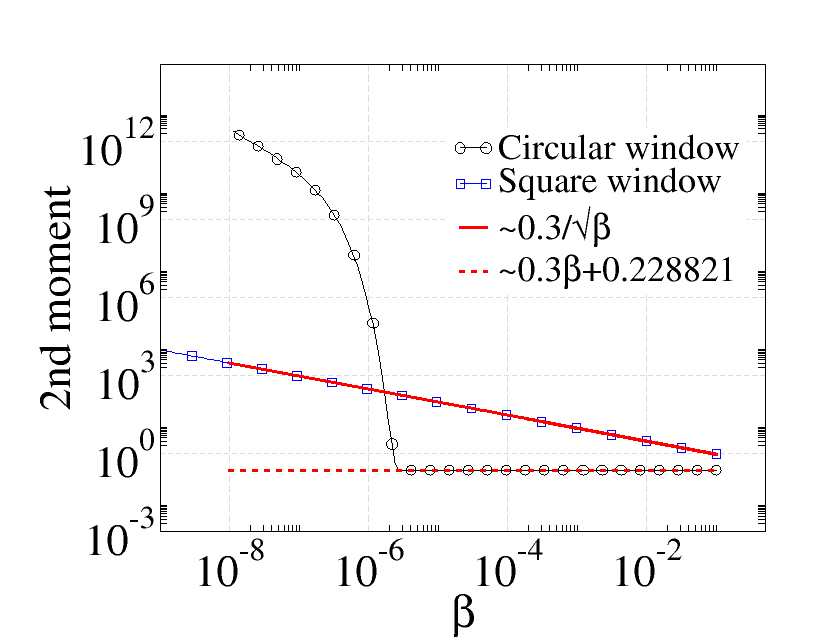}
		\caption{A log-log scale plot of Abelian sums of second moment of total correlation function for circular and square windows as functions of $\beta$.
Black circles and blue squares represent the integrals for circular windows \eqref{rh(r)_convergenceTrick} and for square windows \eqref{xh(r)_convergenceTrick}, respectively.		
To compute \eqref{rh(r)_convergenceTrick}, we summed $Z_k r_k$ up to $r_k =\sqrt{10^7} $, leading the integral to diverge as $\beta \to 0$ in the figure.
We compute the limiting value of the infinite sum \eqref{rh(r)_convergenceTrick} by a linear regression (red dotted line).
For a given $\beta$, we compute the infinite sum \eqref{xh(r)_convergenceTrick} within arbitrarily accuracy, and observe that it diverges in the order of $\beta^{-0.5}$ (red solid line).
 \label{fig:convergence} }		
\end{figure}
In the analysis leading to \eqref{sigma2_expansion}, the convergence of the second moment of total correlation function $\fn{h}{\vect{r}}$ determines the asymptotic behavior of the variance $\sigma_N ^2$ in the limit of $R \to \infty$. 
In order to assign a finite value to \eqref{sigma2_secondTerm} for a circular window, we use the Abelian sum again, yielding
\begin{equation}\label{rh(r)_convergenceTrick}
\fl-\int_{\R^2} r \fn{h}{\vect{r}} \d{\vect{r}} = -\lim_{\beta \to 0^+} \int_{\R^2} r \fn{h}{\vect{r}} e^{-\beta r^2}\d{\vect{r}} = \lim_{\beta \to 0^+} \left[-\sum_{k=1}^\infty Z_k r_k e^{-\beta {r_k}^2} +\frac{1}{2}\left(\frac{\pi}{\beta}  \right)^{3/2}  \right].
\end{equation}
Similarly, the Abelian sum of the second moment of total correlation function for a square window is given by:
\begin{eqnarray}
\fl-\int_{\R^2} \abs{x}\fn{h}{\vect{r}}\d{\vect{r}} &= -\lim_{\beta\to 0^+} \int_{\R^2} \abs{x} \fn{h}{\vect{r}}e^{-\beta r^2} \d{\vect{r}} \nonumber\\
\fl&= \lim_{\beta \to 0^+} \left(-2\sum_{n=1}^\infty ne^{-\beta n^2} \left(1+2\sum_{m=1}^\infty e^{-\beta m^2}  \right)  +\frac{ \sqrt{\pi}}{\beta^{3/2}}\right). \label{xh(r)_convergenceTrick}
\end{eqnarray}
As we can see in \fref{fig:convergence}, the second moment for a circular window \eqref{rh(r)_convergenceTrick} converges to a finite value, while that for square window \eqref{xh(r)_convergenceTrick} diverges.

Figure \ref{fig:convergence} shows the second moments of the total correlation functions for both circular and square windows.
The second moment of $\fn{h}{\vect{r}}$ for circular windows \eqref{rh(r)_convergenceTrick} can be approximated by a straight line, and its $y$-interpolation gives the limit value ($0.228821$), which is consistent with the value given in \cite{HypU_rev}.
The second moment of $\fn{h}{\vect{r}}$ for square window \eqref{xh(r)_convergenceTrick} can be approximated by $0.3/\sqrt{\beta}$, yielding the result
	\begin{equation}
	-\lim_{\beta \to 0^+} \int_{\R^2} \abs{x} \fn{h}{\vect{r}} e^{-\beta r^2}\d{\vect{r}} \sim L.
	\end{equation}

\section*{References}
\providecommand{\newblock}{}

\pagebreak

\begin{center}
\large{\bf Supplementary Material for ``Effect of Window Shape on the Detection of Hyperuniformity via the Local Number Variance"}
\end{center}
\setcounter{equation}{0}
\setcounter{figure}{0}
\setcounter{table}{0}
\setcounter{page}{1}
\makeatletter
\renewcommand{\theequation}{S\arabic{equation}}
\renewcommand{\thefigure}{S\arabic{figure}}
\section{Proof of $\fn{B}{x,y}$}\label{proof of B}
Since the closed expression (50) in the text is continuous and its derivative is piecewise continuous with respect to $x$, we can prove relation (50) by showing that its Fourier series is the same as the infinite sum  \cite{MathPhy1}. 
Since $\fn{B}{x,y}$ is an even function with respect to $x$, we can express it as a cosine series:
	\begin{equation}\label{B_Fourier}
	\fn{B^{\mathrm{(sum)}}}{x,y} = \frac{a_0}{2} +\sum_{k=1}^\infty a_k \cos{\left(2\pi \frac{k}{T_0} x  \right)},
	\end{equation}
	where
	\begin{equation}\label{cosine_coef}
	a_k = \frac{2}{T_0} \int_0 ^{T_0} \fn{B}{x,y}\cos{\left( 2\pi \frac{k}{T_0} x \right)} \d{x},
	\end{equation}
	for non-negative integers $n$. 
Note that the closed expression for $\fn{B}{x,y}$ is the linear interpolation of $\frac{\sin{^2 (2\pi y x)}}{\sin{^2 (\pi y)}}$ whose data points are at $x=k/2$.
Using the expression for the linear interpolation of a function $\fn{f}{x}$ given by
	\begin{equation}
	\fn{f^\mathrm{(1)}}{x} = \sum_{k=-\infty}^\infty \fn{f}{kT}\fn{\Delta_T}{x-kT}	,
	\end{equation}
where $\fn{\Delta_T}{x} = \left(1-\abs{x}/T  \right)\fn{\Theta}{T-\abs{x}}$ and $T$ is the sampling interval,
the closed expression $\fn{B ^\mathrm{(closed)}}{x,y}$ can be written as
	\begin{equation}\label{B_linear interp}
	\fn{B ^\mathrm{(closed)}}{x,y} = \sum_{k=-\infty}^\infty \frac{\sin{^2 (\pi y k)}}{\sin{^2 (\pi y)}} \fn{\Delta_{1/2}}{x-\frac{k}{2}}.
	\end{equation}
Substituting \eqref{B_linear interp} into \eqref{cosine_coef}, for a positive integer $m$, one obtains 
	\begin{equation}\label{a_m step1}
	a_i = \frac{2}{T_0}\sum_{k=-\infty}^\infty \frac{\sin{^2 \left( \pi  \frac{kc}{ T_0} \right)}}{\sin{^2 \left( \pi \frac{c}{T_0} \right)}} \int_0 ^{T_0}  \fn{\Delta_{1/2}}{x-\frac{k}{2}} \cos{\left(2 \pi \frac{i}{T_0} x \right)}\d{x},
	\end{equation}
where $y=c/T_0$, $c$ is a natural number, $T_0 = {L_0}^2$, and $L_0 \equiv \sqrt{n^2+m^2}$ in the text.
Since 
	\begin{equation}\label{Fourier_1D}
	\int_{k/2-1/2} ^{k/2+1/2} \fn{\Delta_{1/2}}{x-\frac{k}{2}} \cos{(2\pi K x)}\d{x} = 2\left( \frac{\sin{(\pi K/2)}}{\pi K} \right)^2 \cos{(\pi K k)},
	\end{equation}
for a real number $K$, \eqref{a_m step1} can be written as
	\begin{eqnarray}
\fl	a_i& = 2\left( \frac{\sin{\left(\pi \frac{i}{2T_0}  \right)}}{\pi\frac{ i}{T_0}\sin{\left(  \pi \frac{c}{T_0}\right)} } \right)^2  \frac{1}{2T_0}\sum_{k=1}^{2T_0} \left[2\cos{\left(\pi k \frac{i}{T_0}  \right)}  - \cos{\left(2\pi k \frac{c+i/2}{T_0}  \right)} -\cos{\left(2\pi k \frac{c-i/2}{T_0}  \right)}\right] \nonumber\\
\fl	& = 2\left( \frac{\sin{\left(\pi \frac{i}{2T_0}  \right)}}{\pi\frac{ i}{T_0}\sin{\left(  \pi \frac{c}{T_0}\right)} } \right)^2 \sum_{k=-\infty}^\infty \left[2 \delta_{i, 2kL_0} -\delta_{i, 2kT_0 - 2c} -\delta_{i, 2kT_0 +2c}\right],\label{a_m step2}	
\end{eqnarray}		
where $\delta_{n,m}$ is a Kronecker delta symbol.
Therefore, non-trivial coefficients $a_m$ are 
	\begin{eqnarray}
	a_0 &=1/\sin{^2 \left( \pi c/T_0 \right)} =1/\sin{^2 \left( \pi y \right)}  \label{a0}\\
	a_{2nL_0 \pm 2c} & = -2 L_0 ^2 /\left( \pi^2 (2n L_0 \pm 2c)^2	\right) = -1/\left( 2\pi^2 (n\pm y)^2	\right) \label{a1} \\
	\end{eqnarray}
Using the identity
	\begin{equation}
	\sum_{k=-\infty}^\infty \frac{1}{\pi^2(k+y)^2} = \frac{1}{\sin{^2 (\pi y)}},
	\end{equation}
and inserting \eqref{a0} and \eqref{a1} into \eqref{B_Fourier}, one can obtain
	\begin{equation}
\fl	\fn{B^{(\mathrm{sum})}}{x,y} = \frac{1}{2\sin{^2 (\pi y)}} - \sum_{n=-\infty}^\infty \left[\frac{\cos{\left( 4\pi (n+y)x \right) }}{2\pi^2 (n+y)^2} \right]
	= \sum_{n=-\infty}^\infty \frac{\sin{^2 \left( 2\pi (n+y)x \right)}}{\pi^2(n+y)^2}. \label{B_f}
	\end{equation}
Note that \eqref{B_f} is identical to the summation form in (50) in the main text, which completes the proof.

\section{Continued fraction and irrational numbers}\label{sec:continued fraction}
\setcounter{equation}{12}
In section \ref{sec:irrational angles} in the text, we heavily used terms in number theory (e.g., badly approximable numbers, $k$-th partial quotients, $\cdots$), but we do not explain them in detail in the middle of the text for the readability.
In this section, we provide definitions of the terms relevant to the rational approximation.
Consider the continued fraction 
\begin{equation}
a_0 +\conFrac{1}{a_1 +\conFrac{1}{a_2 +\conFrac{1}{a_3 +\conFrac{1}{\cdots +\conFrac{1}{a_k}}}}}
\end{equation}
by $\left[a_0, a_1, a_2, \cdots, a_k \right]$ for non-negative integers $a_0,~a_1,\cdots,~a_k$, a real number $\alpha$ can be expressed as
\begin{equation}\label{CF_canonical form}
\alpha = \lim_{n \to \infty} \left[a_0 , a_1, a_2, a_3, \cdots, a_n, \cdots  \right] ,
\end{equation}
where an integer $a_k$ is called $k$-th partial quotient, and is computed by the following recurrence relations:
\begin{eqnarray}
r_0 & =\alpha,~~ a_0 = \floor{r_0} \\
r_k & = \frac{1}{r_{k-1}-a_{k-1}} \\
a_k & =\floor{r_k},
\end{eqnarray}
for positive integers $n$.
If $\alpha$ is a rational number, then its partial quotient is terminated in finite terms. 
Otherwise, its partial quotients form an infinite sequence of non-vanishing integers. 
If we consider the finite number of partial quotients of an irrational number $\alpha$, it gives an rational approximation of $\alpha$, $c_k = \frac{n_k}{m_k} = [a_0, a_1, \cdots, a_k]$ where $c_k$ is called $k$-th convergent of $\alpha$. 
Note that the $k$-th convergent is the best rational approximation of $\alpha$ in the sense that for any rational number $n/m$,
\begin{equation}
 \abs{\alpha -\frac{n}{m}} > \abs{\alpha -\frac{n_k}{m_k}}
 \end{equation} 
as long as $m<m_k$ \cite{Math_Stark_NumberTheory}. 

\begin{table}[h!]
\begin{tabular}{c|c c c c c c c||r|r}
\hline 
$\alpha$ & $a_0$ & $a_1$ & $a_2$ & $a_3$ & $a_4$ & $a_5$ & $a_n$ & $c_\mathrm{double}$ & $c_\mathrm{quadruple}$\\ 
\hline 
$\sqrt{2}$ & 1 & 2 & 2 & 2 & 2 & 2 & 2& $\frac{54608393}{38613965}$ & $\frac{5964153172084899}{4217293152016490}$ \\ 
$\frac{1}{\sqrt{2}}$ & 0 & 1 & 2 & 2 & 2 & 2 & 2& $\frac{38613965}{54608393}$ & $\frac{4217293152016490}{5964153172084899}$ \\ 
$\phi = \frac{1+\sqrt{5}}{2}$ & 1& 1& 1& 1& 1& 1& 1& $\frac{39088169}{24157817}$& $\frac{3416454622906707}{1055742538989025}$\\
\hline
$e$ & 2 & 1 & 2 & 1 & 1 & 4 & \begin{small}$\begin{cond} 2k, &n=3k-1 \\1,&\mathrm{otherwise} \end{cond}$ \end{small}&$\frac{28245729}{10391023}$ & $\frac{2124008553358849}{781379079653017}$ \\ 
$\pi$ & 3 & 7 & 15 & 1 & 292 & 1 & Unknown\footnote{In the canonical form of continued fraction (like \eqref{CF_canonical form}), the general expression for $a_n$ is unknown. Furthermore, whether $\pi$ is a badly approximable number or not remains an open problem \cite{Math_Burger_badlyApproximable}.} & $\frac{80143857}{25510582}$ & $\frac{6134899525417045}{1952799169684491}$ \\ 
$\tanh{(1)}$ & 0 & 1 & 3 & 5 & 7 & 9 & $3n$ & $\frac{2304261}{3025576}$ & $\frac{246256787171955}{323343850850476}$ \\ 
\hline 
\end{tabular} 
\caption{List of partial quotients of some irrational numbers. $c_\mathrm{double}$ and $c_\mathrm{quadruple}$ represent convergents of the values of an irrational number up to 16 and 32 significant figures. Upper three numbers are examples of irrational numbers belong to BAN, and the lower three are not in BAN.
\label{tab:continuedFraction}}
\end{table}
Table \ref{tab:continuedFraction} shows partial quotients of several irrational numbers and their rational approximations up to double and quadruple precision in the form of the irreducible fraction. 
Note that irrational numbers whose partial quotients are bounded, e.g., $\sqrt{N}$ and the golden ratio $\phi$, are ``badly approximable numbers" (BAN) \cite{Math_Burger_badlyApproximable} in the sense that for every rational number $n/m$ 
\begin{equation}
\abs{\alpha-\frac{n}{m}} > \frac{\fn{c}{\alpha}}{m^2}.
\end{equation}
In other words, the best error involving the rational approximation for a badly approximable number decreases like $1/m^2$, while the best error for a non-badly approximable number occasionally can be arbitrary smaller than $1/m^2$.

\section*{References}
\providecommand{\newblock}{}

\end{document}